\documentstyle[12pt,aasms4]{article}

\begin{document}
\def\etal{{\it et al.}}
\small

\title{The Potential-Density Phase Shift Method for Determining\\
the Corotation Radii in Spiral and Barred Galaxies}

\centerline{Xiaolei Zhang}
\centerline{US Naval Research Laboratory} 
\centerline{Remote Sensing Division} 
\centerline{4555 Overlook Ave SW, Washington, DC 20375}
\centerline{}
\centerline{Ronald J. Buta}
\centerline{University of Alabama} 
\centerline{Department of Physics and Astronomy} 
\centerline{514 University Blvd E, Box 870324} 
\centerline{Tuscaloosa, AL 35487} 

\begin{abstract}

We have developed a new method for determining the corotation radii
of density waves in disk galaxies, which makes use of the calculated
radial distribution of an azimuthal phase shift between the potential 
and density wave patterns. The approach originated from
improved theoretical understandings of the
relation between the morphology and kinematics of galaxies,
and on the dynamical interaction between density waves and
the basic-state disk stars which results in the secular evolution
of disk galaxies. In this paper, we present the rationales
behind the method, and the first application of it
to several representative barred and
grand-design spiral galaxies, using near-infrared images to
trace the mass distributions, as well as to calculate the
potential distributions used in the phase shift calculations. 
We compare our results with those from other existing
methods for locating the corotations, and show that
the new method both confirms the previously-established trends of
bar-length dependence on galaxy morphological types,
as well as provides new insights into the possible
extent of bars in disk galaxies. The method also facilitates
the estimation of mass accretion/excretion rates due to bar
and spiral density waves, providing an alternative way of
quantifying the importance of these features in disk galaxies.
A preliminary analysis of
a larger sample shows that the phase shift method is likely to be a 
generally-applicable, accurate, and essentially model-independent 
method for determining the pattern speeds and corotation
radii of single or nested density wave patterns in galaxies.
Other implications of the results of this work include that
most of the nearby bright disk galaxies appear to possess quasi-stationary
spiral modes; that these density wave modes as well as the associated
basic-states of the galactic disks slowly transform over the
time span of a Hubble time due to a collective dissipation process
directly related to the presence of the phase shift between the potential
and density patterns; and that self-consistent N-particle
systems contain physics not revealed by the passive orbit
analysis approaches.
\end{abstract}

\section{INTRODUCTION}

More than two-thirds of all bright galaxies 
in the nearby universe are disk galaxies
(de Vaucouleurs 1963; Buta et al. 1994).
The fraction of disk-dominated galactic systems also increases steadily 
towards
higher redshift. Understanding the kinematics, dynamics and long-term 
evolution
of disk galaxies therefore holds the key to understanding the observed
morphological transformation of galaxies across the Hubble time
(Butcher \& Oemler 1978; Dressler 1980; Koo \& Kron 1992; Ellis 1997;
Couch et al. 1994; Dressler et al. 1994; 
Lilly et al. 1998; Abraham et al. 1999).

The most prominent feature in almost all disk galaxies is the presence
of global patterns such as bars, rings, and spiral structures.  These
patterns have been attributed to density waves which are self-sustained
and long-lasting (Lin \& Shu 1964). Statistical evidence indicates that
most of these patterns can survive for a significant
fraction of a Hubble time (Elmegreen
\& Elmegreen 1983, 1989); N-body simulations of spiral disks had also
substantiated this speculation (Donner \& Thomasson 1994;
Zhang 1998).  These long-lasting density
wave patterns are found to be in fact density wave {\em modes}, formed
by the feedback interactions of oppositely-propagating density wave
{\em trains} in the radial direction, with an amplification mechanism
at the so-called ``corotation resonance" radius
where the density wave pattern and
the underlying differentially rotating disk matter rotate at the
same angular speed (Mark 1976; Toomre 1981). 
For galaxies which have several nested spiral/bar
patterns, it is expected that different pattern speeds and therefore
different corotation radii exist for the different patterns. The 
accurate
determination of the pattern speeds and corotation radii of the density
wave patterns is a key ingredient in determining the dynamical state
of the parent galaxies which will help address their evolutionary 
history
and fate; the pattern speeds of secondary bar patterns also provide
clues on the dark matter distributions in the inner region, and thus
has cosmological implications (Debattista \& Sellwood 2000).

The determination of the spiral density wave pattern speed and
corotation radius has historically been difficult, despite the
great variety of methods proposed in the past to measure either 
parameter. These methods are based on various physical criteria 
or assumptions. The earliest proposal for corotation determination
was based on Lin's (1970) conjecture that an organized two-armed
spiral pattern exists only within the corotation circle, and arms
become more fragmented outside.  Among the first to apply this 
conjecture to physical galaxies were Shu, Stachnik, \& Yost (1971),
who placed the corotation radius where the distribution of
HII regions is seen to end.  Tremaine \& Weinberg (1984, hereafter TW)
developed an approach for pattern speed determination which
was based on the assumption of continuity of a luminosity density
tracer and the measurement of surface brightness and velocity
distributions of the tracer. Canzian (1993) proposed a method for
corotation determination that relied on the qualitative difference
in the appearance of the residual velocity field of a density
tracer inside versus outside
of corotation.  Elmegreen, Elmegreen, \& Seiden (1989) identified
resonances through breaks in spiral arms, whereas Elmegreen,
Wilcots \& Pisano (1998) determined corotation from the transition of
the inward to outward streaming gas velocity. Buta \& Purcell (1998)
and Buta et al. (1998) used the locations of outer rings to infer
the location of corotation in several resonance ring galaxies.
Clemens \& Alexander (2001) used high-resolution
spectroscopic data to find signatures of the pre- and post-shock
gas across the spiral arms, and used this information to determine
the shock velocity and hence the pattern speed.  Oey et al. (2003) used 
an HII region isochrone technique to constrain the corotation radius.  
Various numerical simulation techniques (Hunter et al. 1988;
England 1989; Ball 1992; Lindblad, Lindblad, \&
Athanassoula 1996; Lindblad \& Kristen 1996; Salo et al. 1999; 
Rautiainen, Salo, and Laurikainen 2005) attempted to {\it visually}
match observed galaxy morphology with the simulated density wave
of a given pattern speed, using mostly non-self-consistent (i.e.,
enforced) potential waves. Other methods use offsets between H$\alpha$
and CO arms (Egusa et al. 2006) or between $B$ and $I$-band
spiral arms (Puerari and Dottori 1997) to determine the corotation
radii.  Almost all of the above-mentioned methods have only limited 
ranges of applicability; many lack either accuracy or 
model-independency.

In what follows we describe a novel and general approach to locating 
the corotation radii of density waves in disk galaxies, based on new 
theoretical understandings of the properties of quasi-stationary 
density wave modes (Zhang 1996,1998,1999).  The technique employs 
the use of near-infrared (NIR) images from ground- and space-based surveys
to infer densities and gravitational potentials.  The NIR region of the 
spectrum contains radiation contributions mainly from old giant and 
supergiant stars (Frogel et al. 1996), and is considered a much better 
tracer of the stellar mass distribution than optical bands. 
In the NIR, the effects of dust extinction are also considerably reduced.

The application of new theoretical results of the relation between
the morphological characteristics and the kinematic states of
galaxies provides both a practical approach for corotation
determination, as well as a direct quantitative confrontation
between theoretical predictions and observed galaxy properties.
A further application of the same general approach 
is the estimation the mass accretion/excretion
rates due to bars and spirals. Apart from providing a new way of quantifying
the importance of these features in galaxies (see Buta et al. 2005
and the references therein for other approaches), knowledge of these rates 
is important to our understanding of the secular evolutionary 
transformation of galaxy morphological types (Zhang \& Buta 2006).

\section{POTENTIAL-DENSITY PHASE SHIFT METHOD FOR COROTATION
DETERMINATION}

Our quantitative investigation of the kinematic and dynamical
properties of galaxies using NIR images is
motivated by new theoretical insight on the
role of density waves in the long-term maintenance and evolution of 
galaxies.
Although density wave theory was developed more than
forty years ago (Lin \& Shu 1964), there has been for most of this
time a commonly-held belief that the stellar motion in a
galaxy containing quasi-stationary wave patterns conserves
the Jacobi integral in the rotating frame of the pattern (Binney
\& Tremaine 1987, equation [3-88]). The orbital radii of stars under 
such
{\em applied} potentials generally will not exhibit secular decay
or increase, and there is no wave and basic-state (i.e. the
axisymmetric part of the disk) interaction
except at the wave-particle resonances (Lynden-Bell \& Kalnajs 1972,
hereafter LBK).

It was first demonstrated in Zhang (1996, 1998, 1999) that secular orbital
changes of stars across the entire galaxy disk are in fact possible due
to a collective instability induced by the density wave patterns.
These global patterns were shown to drive a secular energy
and angular momentum exchange process between the disk matter and
the wave pattern, mediated by a local gravitational instability,
or a collisionless shock, at the potential minimum of the pattern
(Zhang 1996). The integral manifestation of this process
is an {\it azimuthal phase shift} between the potential and density spirals,
which results in a secular torque action between the wave pattern and the
underlying disk matter. As a result of the torquing of the wave on the
disk matter, the matter inside the corotation resonance (CR) radius 
loses angular momentum to the wave secularly and sinks inward. The wave 
carries the angular momentum it receives from the inner disk matter 
to the outer disk and deposits it there, causing the matter in the 
outer disk to drift further out.  The resulting secular morphological 
evolution process causes the Hubble type of an average galaxy to evolve 
from late to early (Zhang 1999).

One of the predictions of the above theoretical work is that
for a self-sustained spiral or bar mode, {\it the radial distribution
of the potential-density phase shift should be such that 
it changes sign at the corotation radius} (Zhang 1996, 1998.  Note
that phase shift expression in Zhang 1996 equation [54] formally
diverges exactly at corotation, due to the breakdown of the orbit
approach at the exact resonance, but Zhang 1998 had shown through
the relation of phase shift and angular momentum flux distribution
that for a spontaneously-formed density wave mode the phase shift
crosses zero exactly at corotation).
This happens because the wave rotates more slowly than the disk
stars inside corotation (and thus the wave has negative energy
and angular momentum density with respect to the disk stars),
and vice versa outside corotation. A positive sign of
the phase shift defined as when the potential pattern
lags the density pattern in azimuth, for a radial location inside 
$r_{CR}$, leads on average to an angular momentum loss of the 
population of disk stars in a given annulus inside corotation
(since the phase shift was shown to be directly correlated with
the {\it gradient} of the angular momentum flux being carried outward
by the wave), which further leads to the spontaneous
growth of the wave mode in the linear regime, and the to damping of
the growing wave amplitude towards its quasi-steady-state value in the
nonlinear regime (Zhang 1998).

In the quasi-steady state, the rate of angular momentum exchange
between a skewed density wave pattern 
and the basic state\footnote{The phrase ``basic state'' has been
used to describe the properties of the axisymmetric part of the
disk which forms the boundary conditions for the perturbative
studies of the non-axisymmetric density waves and modes
(see, e.g. Lin \& Lau 1979).  These properties
usually include the radial distribution of surface density
of the disk, the radial distribution of the circular velocity
(which incorporates the influence of any rigid bulge and halo
component), and the radial distribution of velocity dispersion.}
of the disk, per unit area, due to the torquing of the wave
on the disk, is given by
(Zhang 1996, 1998)
\begin{equation}
{\overline{ {{dL} \over {dt}}}} (r) = {\overline {\cal{T}}} (r)
=  - { {1} \over {2 \pi} }
\int_0^{2  \pi}
\Sigma_1(r,\phi)
{ {\partial {{{\cal V}_1}(r,\phi)}} \over {\partial \phi}}
d \phi
,
\end{equation}
where $\Sigma_1$ represents the perturbation density waveform
and ${\cal{V}}_1$ the perturbation potential waveform
(those with the circularly symmetric m=0 component already subtracted).
For two sinusoidal waveforms, eq. (1) can be further written as
\begin{equation}
{\overline{ {{dL} \over {dt}}}} (r)
=
(m/2) A_{\Sigma} (r) A_{\cal{V}} (r) \sin [m \phi_0 (r)]
,
\label{eq:A}
\end{equation}
where $A_{\Sigma}$ and $A_{\cal{V}}$ are the amplitudes
of the density and potential waves, respectively,
$m$ is the number of spiral arms,
and $\phi_0$ is the phase shift between these two
waveforms, defined as being positive if
the potential lags density in the azimuthal direction
in the sense of the galactic rotation.  For most galaxies
the sense of galactic rotation can be determined by
the assumption that the density wave pattern is trailing,
although exceptions do exist which will need to be studied
on a case-by-case basis.

Using eqs. (1) and (2), we see that the radial
distribution of an equivalent azimuthal
phase shift $\phi_0 (r, \phi)$ between the 
(generally nonlinear and nonsinusoidal) 
potential and density patterns 
for spirals or bars in a disk galaxy can be calculated from
\begin{equation}
\phi_0 (r, \phi) \equiv {1 \over m} \sin^{-1}
\left ( {1 \over m}
{{\int_0^{2  \pi}
\Sigma_1
{ {\partial {{{\cal{V}}_1}}} \over {\partial \phi}}
d \phi}
\over
{ \sqrt{\int_0^{2  \pi}
{\cal{V}}_1^2
d \phi}}
\sqrt{\int_0^{2  \pi}
{\Sigma_1}^2
d \phi}  }
\right ),
\label{eq:phi0}
\end{equation}
where the potential $\cal{V}$ is calculated from the Poisson
integral of the density distribution, with the density distribution
itself calculated from the NIR image of a galaxy for an
assumed mass-to-light (M/L) ratio distribution.
The equivalent phase shift is the amount of phase shift
which would be present between two sinusoidal waveforms if each is
endowed with the same energy as the corresponding nonlinear waveform,
and which would lead to the same value for the torque integral 
as would the nonlinear waveforms.
The phase shift is in general non-zero as long as the density pattern
is skewed, i.e., in the case of spirals, twisted or offset bars,
or even some twisted three-dimensional mass distributions as observed
in many high-redshift proto galaxies.

Note that even though these waveforms are given in their perturbational
form (with subscript 1 in the above equations), 
in carrying out the actual calculations we can simply
use the non-perturbed quantities in the numerator, since
the effect of the circularly symmetric components in
$\cal{V}$ and $\Sigma$ will differentiate
and integrate out to zero.  This reduces the sensitivity to 
uncertainties
in the M/L ratio of any axisymmetric mass component 
(or in fact, any non-skewed component, including
triaxial luminous or dark halo component),
including the axisymmetric bulge, halo, and disk components.
The normalization factor in the denominator, however, will be affected by
the axisymmetric components, so these should be subtracted out before
doing the normalization.  However, any uncertainty in the normalization
has no effect on the locations of the positive-to-negative {\em crossings}
of the phase-shift-versus-radius curve, which 
are what we use in the corotation determination.

A practical note for the sign convention of the phase shift is that
we assume the phase shift to be positive when the potential pattern
lags the density pattern in the direction of galactic rotation.
Since the face-on view of the spiral (or twisted bar)
disk comes in two flavors, i.e., either the S-sensed pattern or the
Z-sensed pattern, these in general indicate a counter-clockwise
or a clockwise rotation direction, respectively, if the pattern
is assumed to be trailing.  In the above definitions, therefore,
the sense of the phase shift is correct only for the S-sensed 
spiral or bar (which winds and rotates in the same counter-clockwise
direction as used in the azimuthal angle definition).  
For a Z-sensed pattern, the calculated phase shift
curve should be negated to obtain the correct sense of the
phase shift in the direction of galactic rotation.
In practice, we usually generate both versions (called a
``plus'' plot and a ``minus'' plot) for the phase-shift-versus-radius 
distributions, and compare both with the image
of the galaxy, since in certain cases the sense of the winding
is not so obvious (for some barred galaxies where the bars
are fairly straight, or for cases where leading arms may
be present, as for the case of NGC 4622 which we discuss
in \S 3.4).

After obtaining the phase-shift-versus-galactic-radius plot,
the positions of the successive corotation radii can be
read off as the positive-to-negative crossings of the phase shift
plot. These mark the locations where the direction of angular
momentum flow between the disk matter and the density wave
changes sign, at the quasi-steady state of the wave.
The negative-to-positive crossings of the phase shift, on
the other hand, generally mark the transition locations of nested
modal structures, i.e., each such negative-to-positive
crossing is likely to be where the pattern speeds of the
density waves change abruptly.  This is because the density
wave should have positive angular momentum inside corotation
and negative angular momentum outside corotation.  The 
negative-to-positive
crossings mark the boundaries where the wave deposit angular momentum
onto the disk (for the inner mode outside its corotation) and the
wave takes away angular momentum (for the outer mode inside its own
corotation).
So the phase shift distribution provides not only
a way for the objective determination of density
wave corotation radii, but also for the objective
determination of the radial extent of the individual
modes in the nested-modes cases, i.e., what are the
effective radii where adjacent modes decouple: something which
methods such as TW could not determine in a model-independent
way.  This assertion of course is based on the assumption
that the radial
penetration of the different sets of modes are minimal,
which should be expected if the morphology of the modes
are quasi-steady, since the inter-penetration of modes
with different pattern speeds will result in rapidly-changing
overall density wave morphology.

We emphasize that for the above approach to work, {\it the individual 
density
wave modes in the galaxies of interest have to have
already achieved quasi-stationary state,}  
for otherwise one would not expect the purely morphological
characteristics (i.e. an image of a galaxy) to correlate with
kinematic-status information (i.e. the rotational state
of the disk stars and density waves).  Therefore, a demonstration
of the validity of this approach, i.e., if the corotations we 
determined
actually correspond to visible resonant features in the images
of galaxies, also confirms the quasi-steady state of the density
wave modes in the relevant galaxies.

\section{APPLICATION TO INDIVIDUAL GALAXIES}

The potential-density phase shift, as well as its sign change across corotation,
has been confirmed with N-body simulations (Zhang 1996, 1998), which verified
that the method gives the same corotation radius as that derived from
the pattern speed of the density wave in conjunction with a galactic rotation 
curve. Our present study represents the first detection of the phase shift in 
observed galaxies, based mainly on $H$-band (1.65$\mu$m) images from the
Ohio State University Bright Galaxy Survey (OSUBGS, Eskridge et al.  2002) 
as well as on other smaller samples.  

NIR images can be used to measure the phase shifts because, as previously 
noted, such images trace the stellar 
mass distribution better than do optical images, and for this reason
are also the best to use for calculating gravitational potentials (e.g, 
Quillen, Frogel, \& Gonz\'alez 1994).  
For most of the calculations
given below, the M/L ratio has been assumed to be a constant
independent of radius unless otherwise noted. The exact value of this
constant is not important, since it cancels out
between the numerator and the
denominator in the definition equation 3.  The validity
of this assumption is further justified in \S 3.2.
Following Quillen et al.,
we calculate the two-dimensional (2D) 
potential at the plane of the galaxy from

\begin{equation}
\Phi(x, y, z) = -G \int \Sigma(x', y') g(x-x', y-y') dx' dy'
\end{equation}
where the 2D Green's function $g(r)=g(\sqrt{x^2 + y^2})$ is given by
\begin{equation}
g(r) = \int_{-\infty}^{\infty} {{\rho(z) dz} \over {\sqrt{r^2 + z^2}}}
,
\end{equation}
where $\rho_z (z)$ is the normalized z distribution of the volume 
density
of matter, assumed to be independent of galactic radii r.  In practice,
we found that the several different forms of the
assumed $\rho_z (z) $ distribution given in Quillen et al. (1994) do
not give significantly different phase shift results, especially
for the zero crossings, even though they
do change the absolute values of the calculated potential by about 
20\%.  
This insensitivity to the exact z profile of the density distribution
is related to the fact that phase shift is determined mostly by
the global distribution of the pattern pitch angle
and radial density variation, and is insensitive to how ``puffed up''
the pattern is in the z-direction.  We have available Quillen's original
code for potential calculation written in C.  But since that code assumed
arbitrary units, which was fine for the phase-shift calculation,
but not adequate for the mass-accretion calculation which we are
also interested (see \S 4.3 of this paper),
we had completely re-written the code in Fortran,
and adopted also a different FFT routine.  The resulting code gives
potential in physical units, and also consistently out-performs
Quillen's original code
in terms of resolving the central structures of galaxies in the
phase shift results, which turned out to be important for galaxies
possessing nested density wave modes, such as in the cases of
NGC 1530 and NGC 4321.  An example of the calculated potential 
for NGC 1530 using the above approach will be given in \S 3.2.

After the surface density map and the potential map for each galaxy 
are thus determined (on the cartesian grid), 
the phase-shift-versus-galactic-radius curve 
is calculated using equation \ref{eq:phi0}
on an exponential polar grid similar to that used in N-body simulations
(Zhang 1996), with values of the potential and density on the polar
grid obtained
through an interpolation between the polar and cartesian grid.
The azimuthal distributions for the potential and density at each
radius on the polar grid
are first fitted with cubic-spline curves, the differentiation
and integration are then performed based on the fitted potential
and density curves using routines in Numerical
Recipes (Press et al. 1992). For most galaxies the convergence
of the integration is rapid. Occasionally, however, when the
phase-shift-versus-radius curve is particularly noisy there exist
some radii where the phase shift calculations fail to converge
after a reasonable number of iterations (so far this happens for
less than 1\% of the galaxies we have calculated). 
Since this occurs for isolated radii only
we used interpolation of the phase shift results of neighboring points
to arrive at the values for these radial locations.

In the rest of this section, we will describe representative results 
of the phase shift calculation and corotation 
determination for a number of galaxies we have analyzed so far,
and defer the description of such calculations for the majority of the
galaxies in the OSUBGS sample to a sequel paper (Buta \& Zhang 2007).
The detailed assessment of the accuracy of the method, including
the dependence of the results on the uncertainties in vertical scale height,
on the extent of the galaxy image size used, on the signal-to-noise ratio
of the image, on the wave band of the image used, as well as on the 
uncertainties in the orientation parameters used for the deprojection 
of the image will also be given there.

\subsection{Representative Cases}

Among the galaxies we have analyzed so far,
the simplest cases to interpret (and often also the most striking
in appearance) are those where the phase-shift-versus-galactic-radius curve 
has only a single major positive-to-negative
crossing across the entire disk, though some may contain
a second, or even a third, minor crossing in the central region of the galaxy
as well, signaling secondary, nested nuclear spiral or bar patterns. 
Examples from the OSUBGS which fall into this class
include NGC 4051, 4314, 4548, 4665, 5054,
5247, 5248, 6215, 7479, and 7552.

The galaxies listed above can be further divided into two
sub-categories: (1) those 
with prominent bars, such as NGC 4314, 4548, and 7479, which have
a derived CR circle that lies very near to (or slightly larger than)
the end of the bar;  (2) those containing grand design spirals such as
NGC 5054, 5247, and 5248, where a single CR circle lies in the middle 
(or sometimes near the end) of the main spiral pattern.

For the kind of bars where $r_{CR}$ is at the end of the bar and there
is spiral pattern emanating from the bar end, the inner 
and outer patterns are most likely corotating, that is to say, in the case 
where the bar ends at its $r_{CR}$ we most likely have a bar-driven spiral
outside of the bar. For the case of a spiral pattern with a single $r_{CR}$
ending in the middle of the 
spiral pattern, the spiral is likely to have extended 
all the way to its outer Lindblad resonance (OLR).

\begin{figure}[ht!]
\plotone{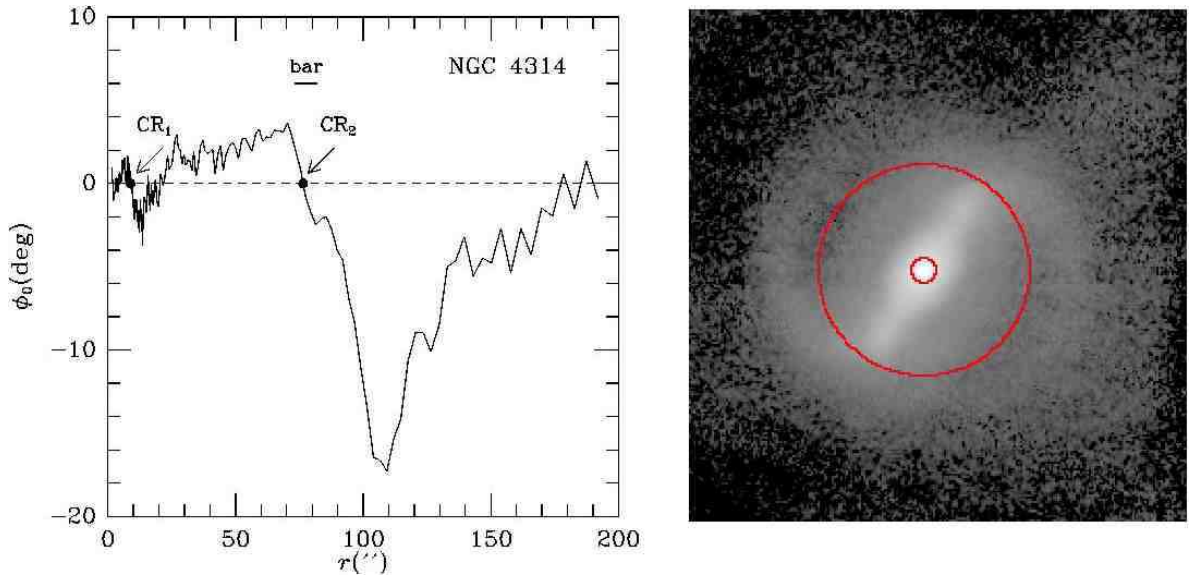}
\caption{{\it Left}: Calculated phase shift versus radius for OSUBGS 
galaxy NGC 4314, type SBa. The arrows point to the locations of two 
positive-to-negative phase shift crossings 
(radii 9\rlap{.}$^{\prime\prime}$0 and 76\rlap{.}$^{\prime\prime}5$), 
interpreted as CR radii of inner and
outer patterns. The short horizontal line indicates the radial range of 
the
bar obtained based on two definitions given \S 3.5. 
{\it Right}: Deprojected near-infrared $H$-band (1.65 $\mu$m)
image of NGC 4314 in log scale,
with the corotations determined by the phase shift method superimposed 
as
circles.}
\label{fg:fg4314}
\end{figure}

In Figure \ref{fg:fg4314}, left frame, we show the calculated 
phase-shift-versus-galactic-radius plot for the nearly face-on SBa
galaxy NGC 4314, based on a deprojected OSUBGS $H$-band image.
The gravitational potential has been calculated under the assumption
of a constant M/L ratio, and an exponential
vertical density distribution with scale height
$h_z$ of 1/4 the radial scale length $h_R$ (de Grijs 1998; Laurikainen 
et al. 2004), using the approach described before. 
The location of the corotations (two in this case) are marked by
the positive-to-negative crossings of the phase shift curve ({\it arrows}) 
at 9\rlap{.}$^{\prime\prime}$0 and 76\rlap{.}$^{\prime\prime}$5.
In Figure \ref{fg:fg4314}, right frame, these two corotation radii are 
plotted on the same $H$-band image. 

As can be seen from the image for NGC 4314 overlayed with the
corotation circles, the outer corotation corresponds to the end of the bar.  
This outer corotation radius determined by the phase shift method
is found to be very close to the value determined for this galaxy by
Quillen et al (1994).  The inner
corotation indicates the existence of an inner density wave pattern,
likely associated with the well-known nuclear ring/spiral in this
galaxy (Sandage 1961). This feature has a radius of 8$^{\prime\prime}$
according to Buta \& Crocker (1993). Note that even though 
the inner density wave pattern is not well resolved due to the 
limited spatial resolution of this $H$-band image,
a potential-density phase shift cross-over is nonetheless clearly
predicted.  This apparent ``super-resolution'' effect is
achieved because the potential is obtained as the Poisson integral
of the global density distribution of the entire disk matter, and is 
not
determined solely by the {\em local} density distribution near the
central region of the galaxy.

\begin{figure}[ht!]
\plotone{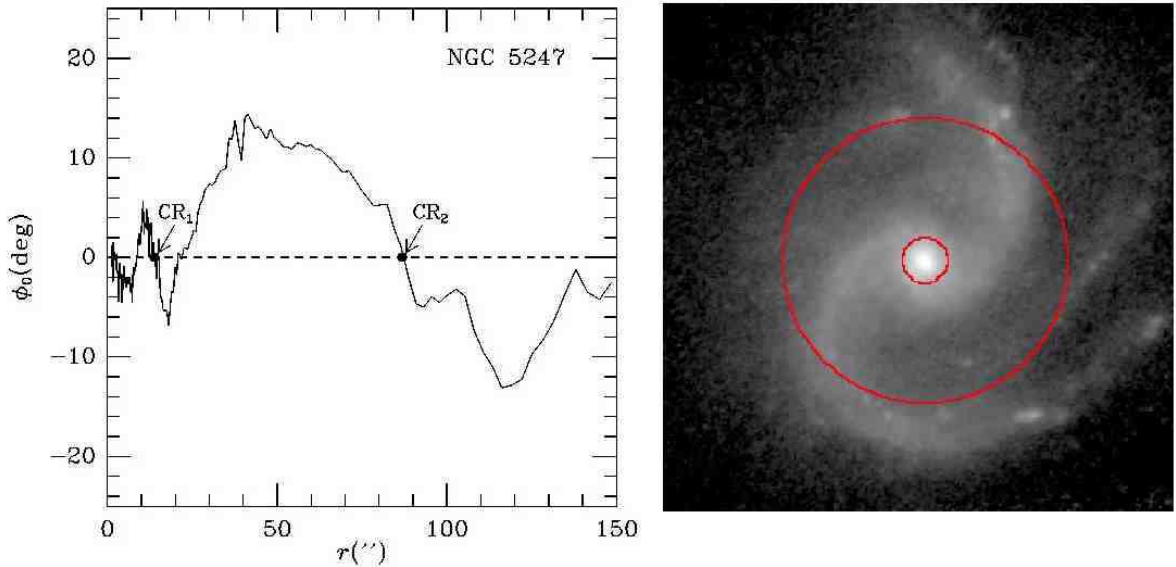}
\caption{{\it Left}: Phase shift versus radius for the 
ordinary spiral NGC 5247, 
showing two major positive-to-negative crossings at 
r=87$^{\prime\prime}$ and at $r$=14$^{\prime\prime}$. 
{\it Right}:
Deprojected near-infrared $H$-band (1.65 $\mu$m)
image of NGC 5247 in log scale,
with the corotations determined by the phase shift method superimposed 
as
circles.}
\label{fg:fgspiral}
\end{figure}

In Figure \ref{fg:fgspiral} we show the classic case of an ordinary
grand-design spiral pattern NGC 5247, and the corotation circles 
determined through the phase shift method.  The outer crossing, 
which indicates the location of the main CR circle, is in 
the middle of the outer spiral arms.  The outer corotation
radius we determined
appears to be close to where the inner dust lanes 
truncate on the two spiral arms (this effect is more obvious from a
B-band image of this galaxy), which is usually a good and
reliable indication of the corotation radius.

Contopoulos \& Grosbol (1996,1988) had previously determined the corotation
radius for this galaxy under the assumption that the 4:1 resonance (where
$\Omega - \kappa/4 = \Omega_p$, with $\kappa$ the epicycle frequency and
$\Omega$ the angular speed of the disk matter, and $\Omega_p$ the 
pattern speed of the density wave) lies 
near the end of the strong spiral pattern.
Their analysis placed this resonance at 12 kpc, which corresponds
to 116$^{\prime\prime}$ assuming a distance of 21.3 Mpc. The corresponding
location of the corotation radius from their analysis is 23 kpc, 
or 222$^{\prime\prime}$, well outside the NIR image of the spiral pattern.
We note that Contopoulos \& Grosbol's original analysis was
based on matching the surface density obtained from
passive orbital responses calculated under a forced
potential to that of the given galaxy morphology. 
They had also assumed that the sharp truncation of the spiral pattern 
occurs near the 4:1 resonance, as the starting point
of their model construction process.  We will present extensive
arguments later on (in \S 4.1) of why the passive orbit approach
for studying the extent of density waves in galaxies
has limitations when collective dissipation sets in, as is the
case for the density wave modes in physical galaxies.

\subsection{Nested Resonances, and Effect of M/L Ratio}

NGC 1530 is a high luminosity SB(rs)b spiral with one of the
strongest known bars. We use a $K_s$-band
(2.15 $\mu$m) image of this galaxy (Block et al. 2004) to 
conduct some more detailed analysis of the nested resonance
features already encountered for the
two galaxies in \S 3.1.  We will also use this galaxy as an example
to illustrate the effect of M/L ratio (and its uncertainties)
on the phase shift and corotation determination, as well as
on determining the mass flow rate (accretion/excretion)
as a function of radius (\S 4.3). A detailed image
of the calculated potential for this galaxy is given
as a representive example, using the numerical codes we had described 
at the beginning of \S 3.

We examine
M/L ratio variations using the color-dependent formulae of
Bell and de Jong (2001). 
The conversion of the $K_s$-band image of NGC 1530 into
a surface mass density map was accomplished using a 
$V-K_s$ color index map. The $V$-band image was obtained with
the Nordic 2.5m Optical Telescope, while
the $K_s$-band image was obtained with the 4.2m William
Herschel Telescope. These images were matched in scale
and orientation using IRAF routines GEOMAP and GEOTRAN.
Zero points were obtained using published photoelectric
aperture photometry for the $V$-band image, and the 2MASS
$K_s$-band magnitude in a 14$^{\prime\prime}$ aperture provided
on the NED website. Each image was cleaned of foreground
stars, deprojected using a position
angle of 8$^{\circ}$ and inclination of 45$^{\circ}$ from
Regan et al. (1996), and rotated so that the bar is
horizontal. The resulting $V-K_s$ color index map showed
strong red dust lanes in the bar and bluer spiral arms.
Assuming a reference bar color of $(V-K_s)_{ref}$=3.47 
(corrected for Galactic extinction as given by NED [Schlegel
et al. 1998]), we were
able to effectively correct the color index map
for the dust lanes using a simple screen approximation\footnote{The 
reference bar color refers to the reddening-free color.
Any bar point redder than this is assumed to be
reddened. Any galaxy point bluer is assumed to have that color 
as intrinsic. This may not be correct in every case, but 
no other large-scale reddening in NGC 1530 is as serious 
as that in the bar dust 
lanes. The screen approximation assumes all the dust doing 
the reddening is in the foreground of the stars being
reddened, rather than mixed with the stars. Thus, the
dust acts like a ``screen." This is not perfectly accurate, but
it worked well enough for our analysis.}. 

\begin{figure}[ht]
\centerline{
\plottwo{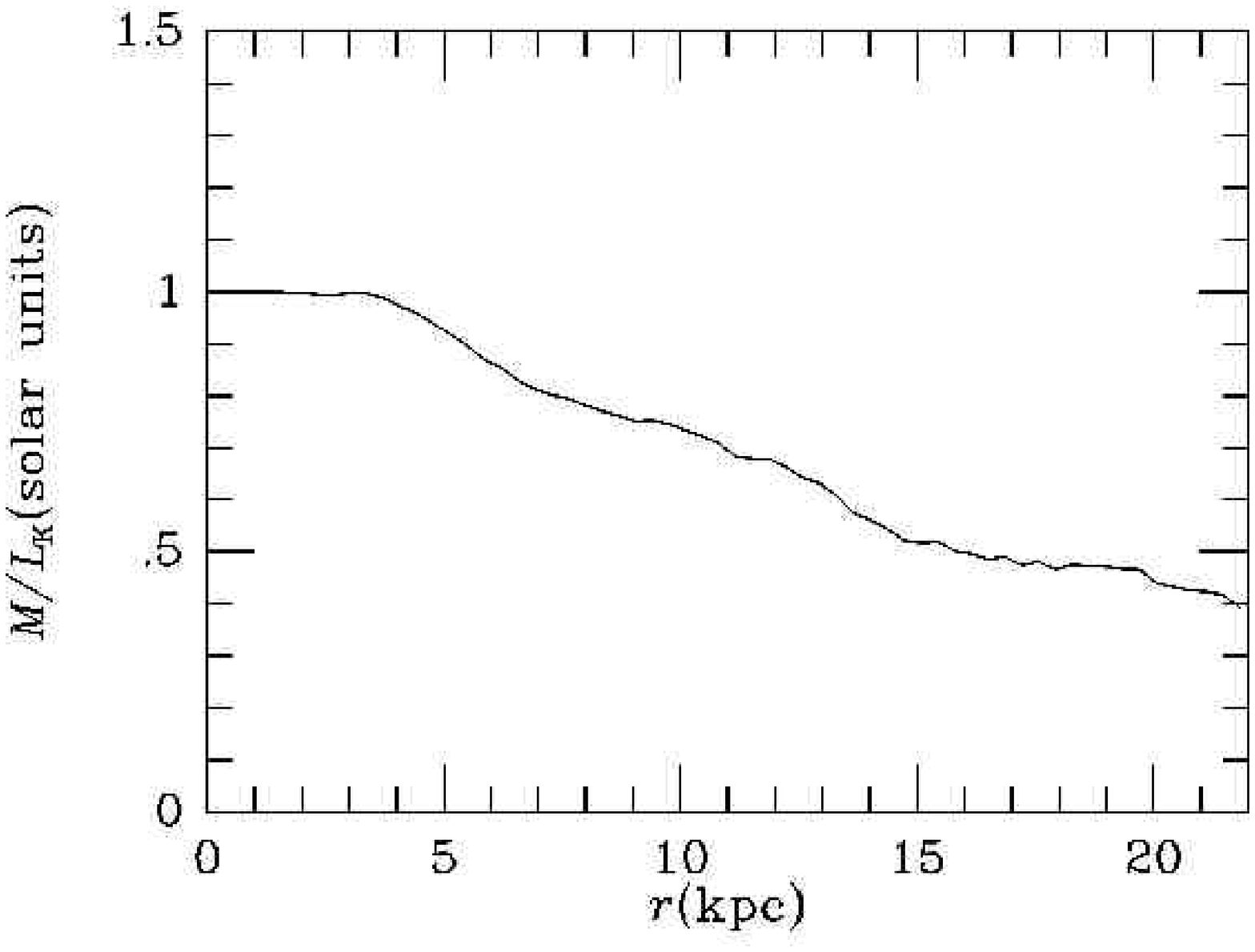}{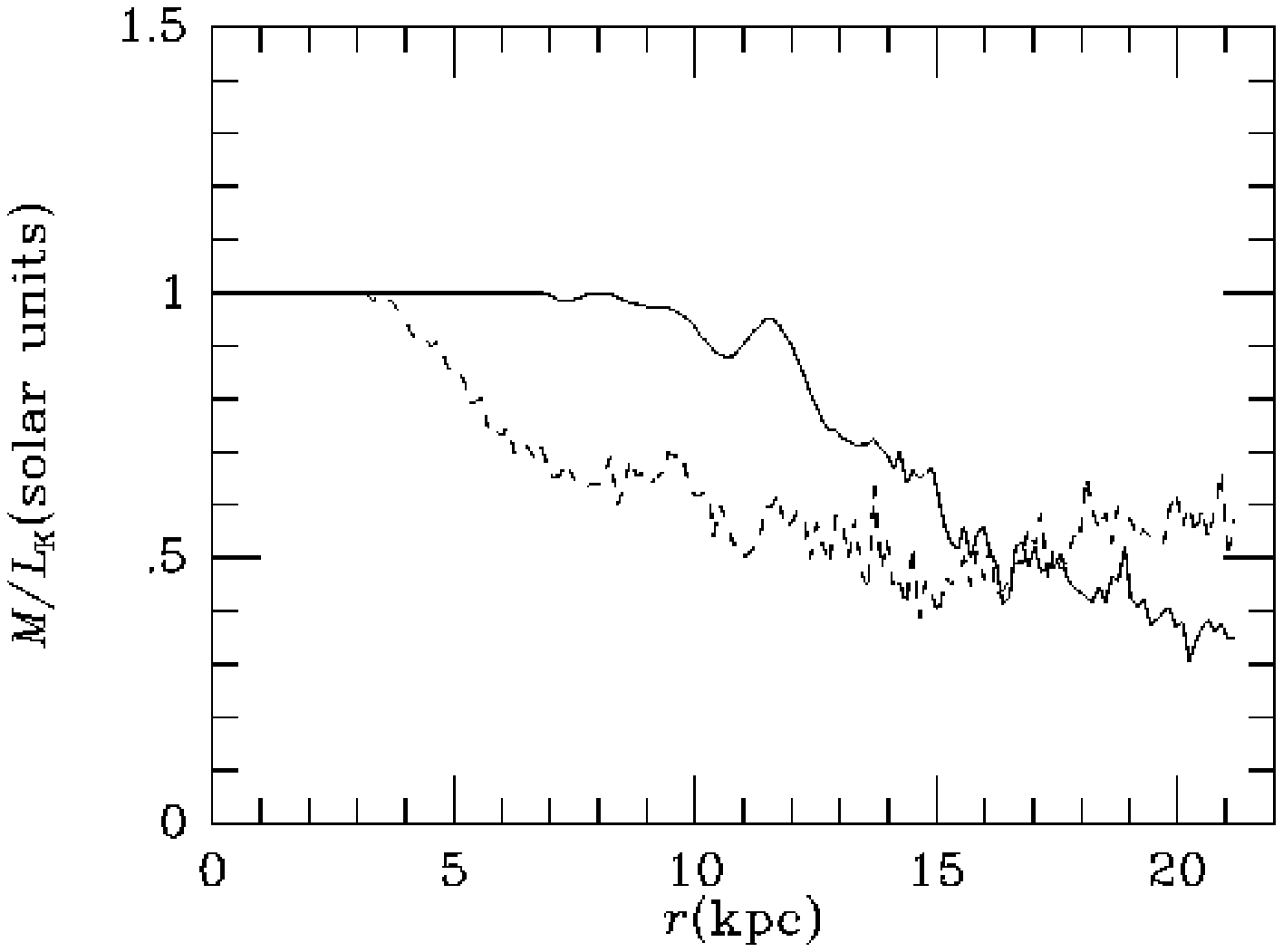}
}
\caption{{\it Left}: Radial dependence of the 
azimuthally-averaged M/L ratio for
NGC 1530. {\it Right}: Interpolated 
M/L profile along the bar (solid curve)
and 90 degrees from the bar (dashed curve).}
\label{fg:fg1530mol}
\end{figure}

\begin{figure}[ht!]
\plotone{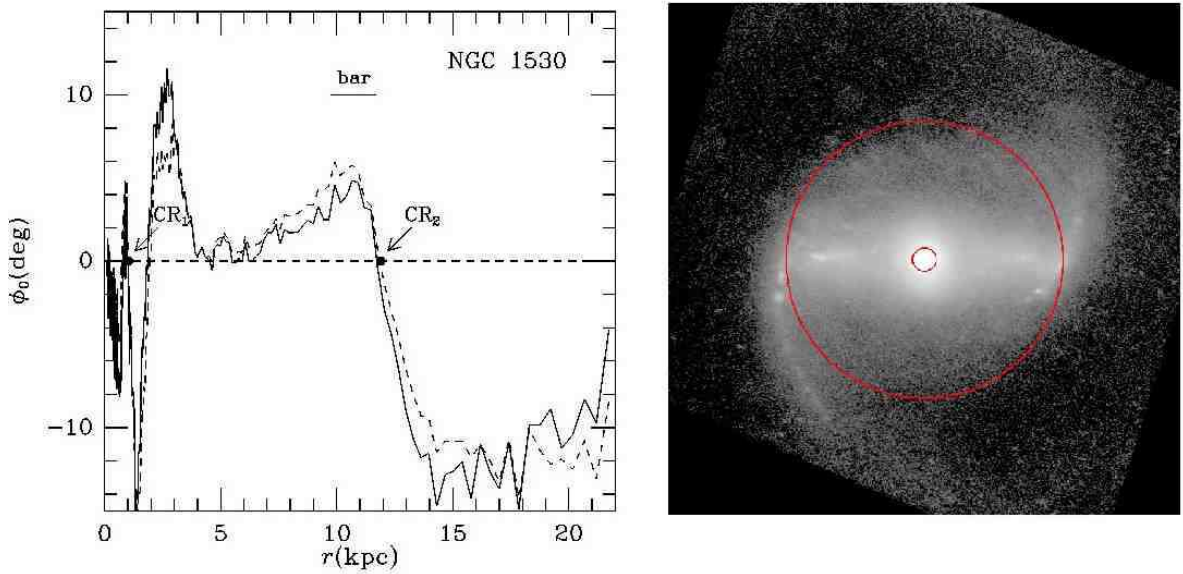}
\caption{{\it Left}:
Phase shift versus galactic radius calculated for
the SBb galaxy NGC 1530 from a $K_s$-band image.
The solid curve was calculated using a position-dependent M/L
map, and the dashed curve was
calculated using the surface density map assuming cosntatn M/L
ratio, Two CR circles are indicated.
The short horizontal line indicates the range for the end of the
bar as determined based on the two definitions given in \S 3.5.
{\it Right}: 
Deprojected $K_s$-band image (Block et al. 2004)
of NGC 1530 in log scale, with superposed
corotation circles, rotated such that the bar is horizontal. 
This image has been corrected for stellar M/L ratio 
variations following Bell \& de Jong (2001). The radius scale
is based on a distance of 36.6 Mpc from Tully (1988).}
\label{fg:fg1530}
\end{figure}

The corrected color index map was then used to derive
a M/L ratio map using $log(M/L)=-1.087+0.314(V-K)$
from Table 1 of Bell \& de Jong (2001), assuming $V-K$ $\approx$ 
$V-K_s$.
This is for the ``formation epoch model with bursts" that
uses a scaled Salpeter initial mass function, and it
gives $M/L$=1.0 for the reference color of 3.47.
Using the NED zero point, the $K_s$-band image was converted
to solar $K$-luminosities per square parsec using an absolute
magnitude $M_K(\odot)$=3.33 from Worthey (1994). Multiplying
the original image
by the $M/L$ map gave the surface mass density image.
The main effect of the $M/L$ correction was to reduce the surface
density of the spiral compared to the bar, thereby weakening 
the effect of the spiral.
The azimuthally-averaged M/L ratio for NGC 1530 is given in Figure
\ref{fg:fg1530mol} (Left), along with the break-down of the
M/L along the bar and 90 degrees from the bar in Figure 
\ref{fg:fg1530mol}
(Right).  This procedure gave the surface mass density map in Figure 
\ref{fg:fg1530} (Right), which is in units of $M_{\odot}$ pc$^{-2}$. 

Figure \ref{fg:fg1530}, left frame, shows the calculated
phase shift versus radius for NGC 1530, using either a constant M/L
ratio (dashed curve) or a position-dependent M/L ratio (solid curve).
We observe that the effect of varying the M/L ratio
changed mostly the scalings of the phase shift results,
and the zero crossings changed by only about 1\%.  
This is to be expected from the arguments
given in \S 2, which showed that the radial M/L ratio change for
the axisymmetric components does not impact the values of the
zero-crossings of the phase shift, and much of the M/L variation
in physical galaxies is due to the M/L change of the axisymmetric
components only, since the population of stars supporting
the density wave is relatively homogeneous in composition
(i.e. mostly old giant and supergiant stars).  The right frame of
Figure \ref{fg:fg1530} displays the
converted surface density map of the galaxy
image in log scale, with two superimposed corotation 
circles determined using the phase shift method ({\it arrows}).
The radial scalings used for NGC 1530 in the above
calculations are based on a distance of 36.6 Mpc 
from Tully (1988). 

\begin{figure}[ht!]
\vspace{3.6in}
\centerline{
\includegraphics{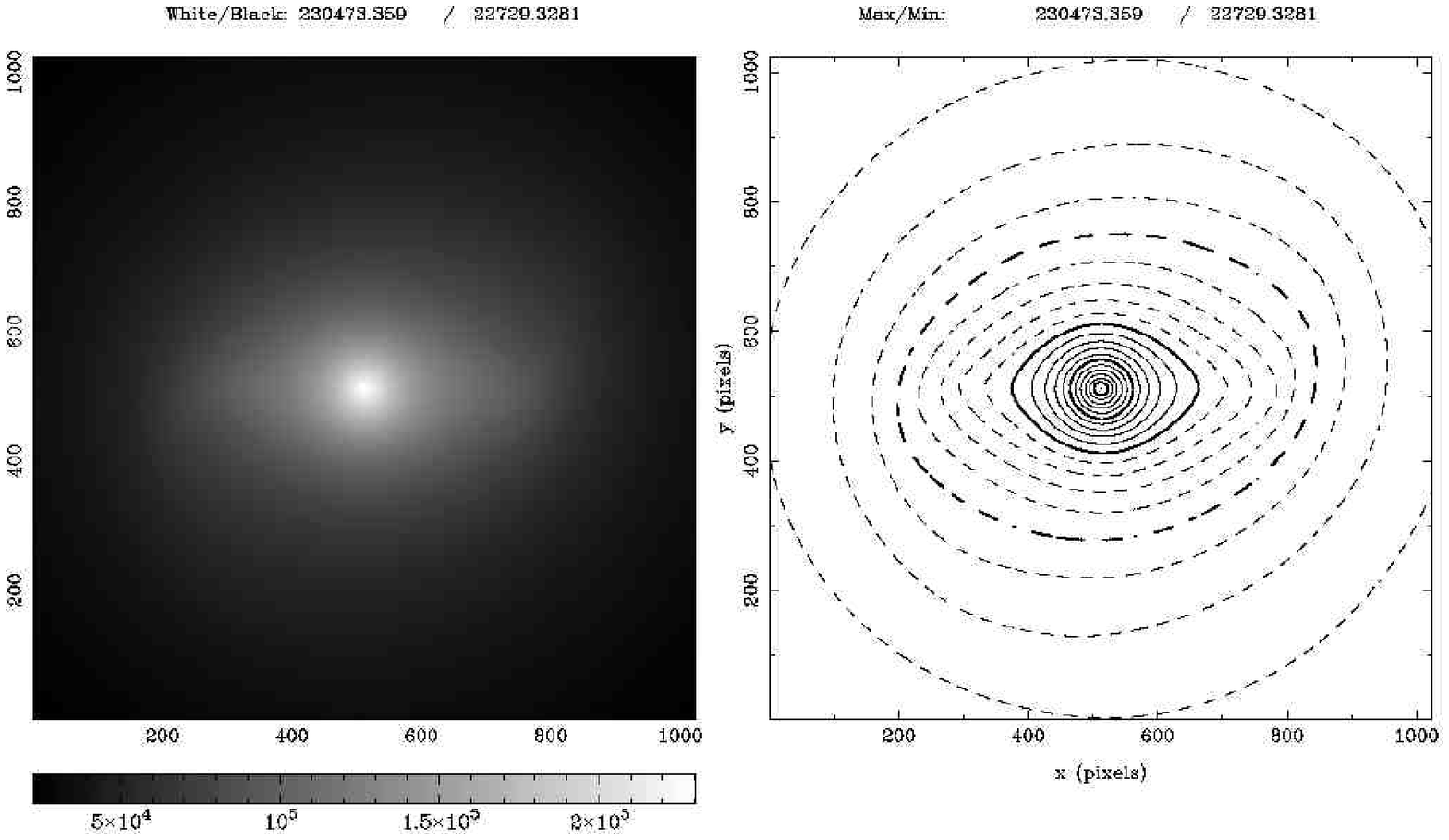}
}
\caption{Calculated potential distribution for NGC 1530 in grey
scale (Left) and in contours (Right). The map unit is
$(km/s)^2$.}
\label{fg:fgN1530pot}
\bigskip
\end{figure}

For reference, we also present in
Figure \ref{fg:fgN1530pot} an image of the calculated potential
distribution used in the phase shift calculations for this
galaxy.  From this figure one can discern the slow isophotal
twist, with a level of skewness that is more gradual than the 
density pattern skewness observed in Figure \ref{fg:fg1530} (right),
consistent with the sense of phase-shift-sign-change inside
and outside corotation (which would mean that a potential
pattern is in general {\em straighter} than the density pattern).

As was the case 
for NGC 4314, the inner CR for NGC 1530 is associated with a small
central pattern with size about 1.1 kpc, and the overall
extent of this central feature appears to be delineated by the
negative-to-positive crossing on the phase shift plot.
Immediately following the central pattern is the intermediate
oval pattern between 1 and 2 kpc, with its own CR at about 1.3 kpc
(this was marked as CR$_1$ in this plot since the inner-most corotation
radius is not very well resolved by the current calculation).
The main CR encircles the ends of the
bar at a radius about 13 kpc, with a pair of
bar-driven spiral arms emanating from the
end of the bar outside of the main CR.

\begin{figure}[ht!]
\bigskip
\plottwo{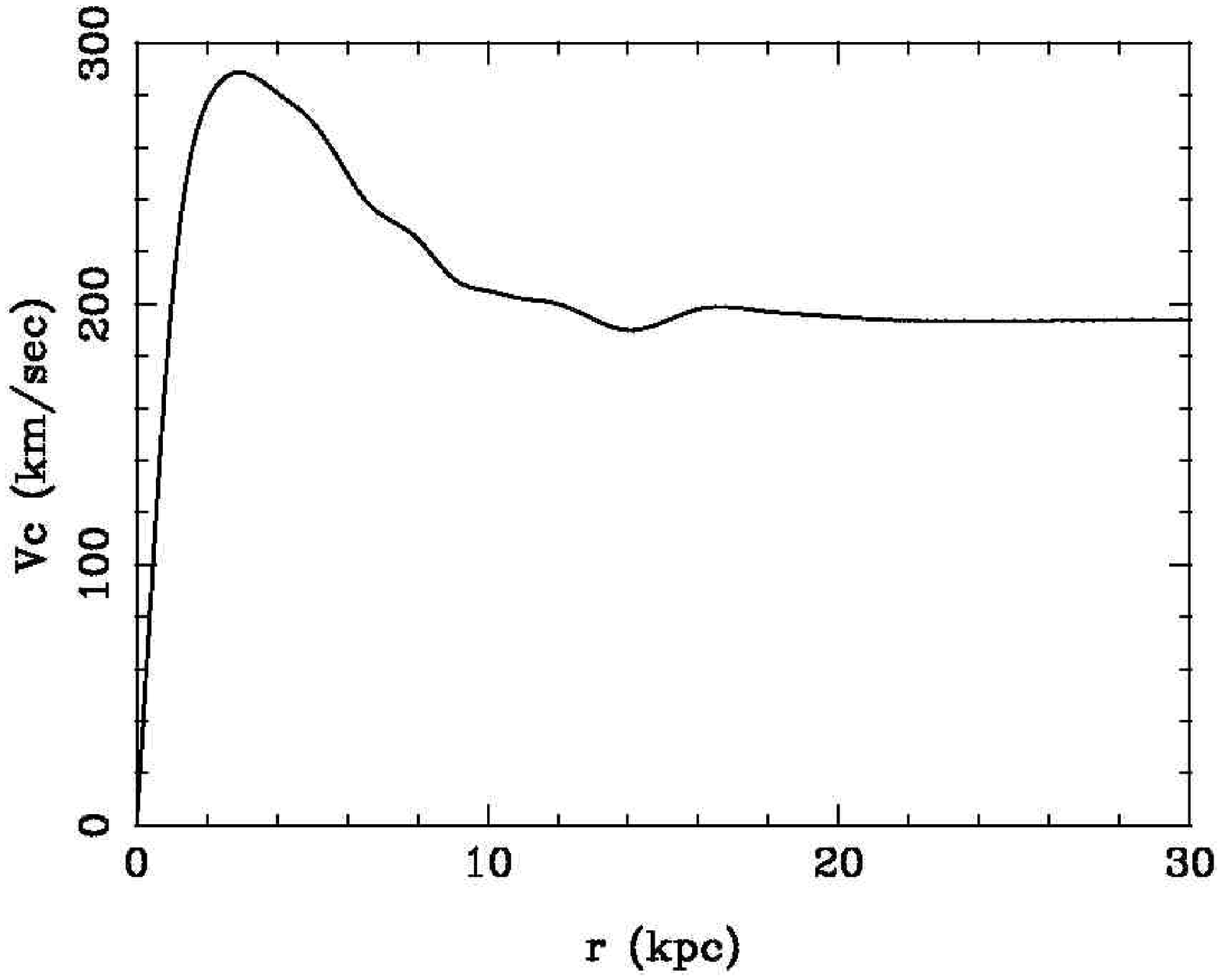}{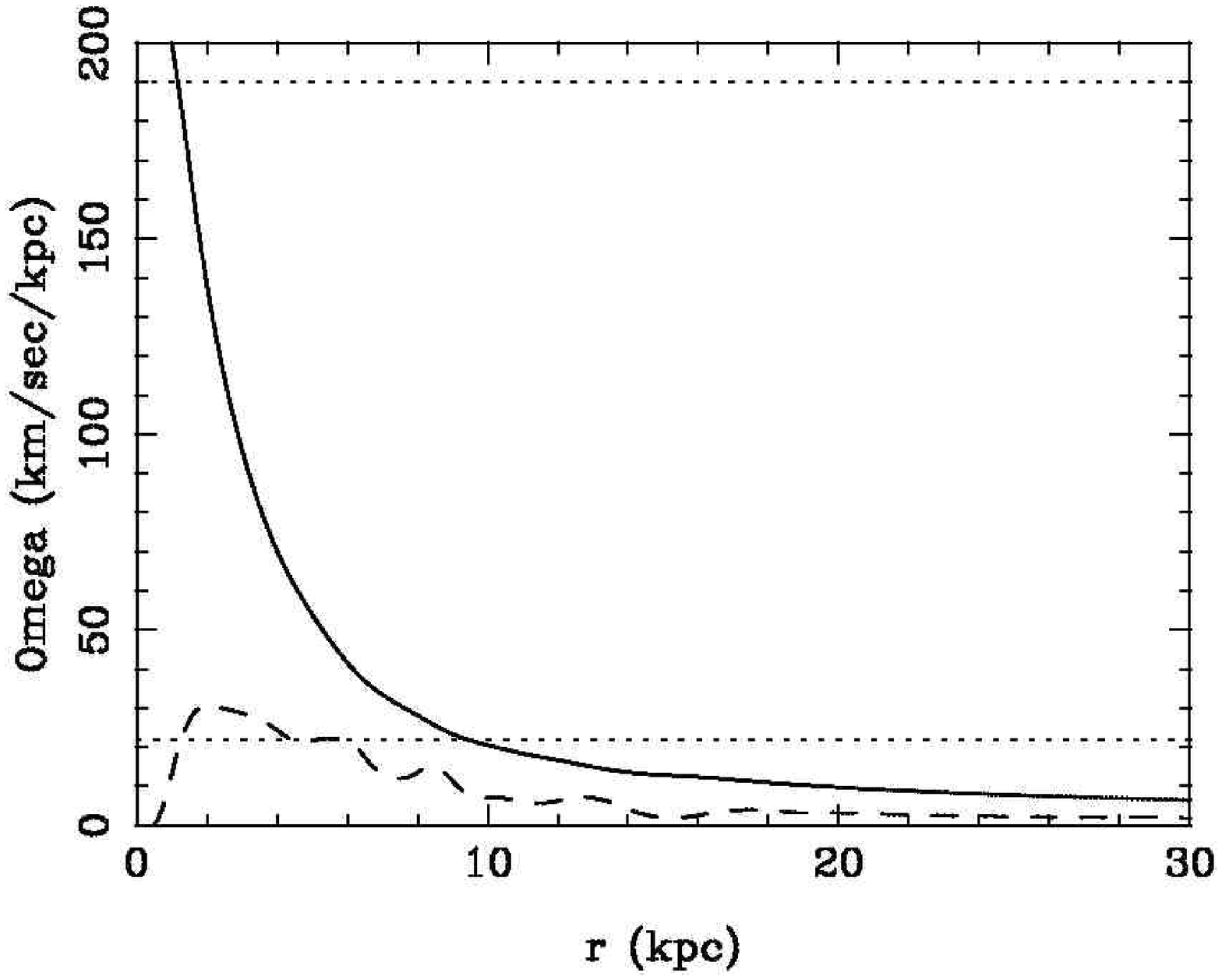}
\caption{{\it: Left:} The rotation curve of NGC 1530
(Regan et al. 1996).
{\it Right:} Plot showing the quantities $\Omega$ (solid curve), 
$\Omega_p$
(dotted curves) and $\Omega - \kappa/2$ (dashed curve).}
\label{fg:fg1530res}
\end{figure}

Figure \ref{fg:fg1530res} uses published kinematic information to
calculate and plot several kinematic diagnostic curves for this galaxy,
which we use to compare with the nested resonance features 
determined
using the phase shift method.  The left frame shows the rotation curve
from Regan et al. (1996), based on the atomic and ionized gas kinematics,
while the right frame shows the calculated angular speed $\Omega$, 
$\Omega-\kappa/2$, and the two pattern speeds determined by the phase 
shift method.  We see from the frequency curves that
the inner-inner-Lindblad resonance (IILR) of the outer pattern
(signified by the {\it inner} intersection of $\Omega_p$ of the 
outer pattern, i.e., the lower dotted curve, with the $\Omega-\kappa/2$ curve, 
or the dashed curve, at about 1.3 kpc radius),
is simultaneously the corotation radius of the inner nested secondary
pattern (signified by the intersection of the $\Omega_p$ of the inner
pattern, i.e., the higher dotted curve, with the $\Omega$ curve, or the
solid line).  This kind of nested resonance coupling 
has been found for other observed galaxies
(Zhang, Wright \& Alexander 1993; Erwin \& Sparke 1999; Laine et al.  2002 ),
and in N-body simulations (Schwarz 1984; Zhang 1998; Rautiainen \& Salo 1999),
and varying theoretical mechanisms had been proposed for its interpretation
(Contopoulos \& Mertzanides 1977; Tagger et al. 1987; Friedli \& Martinet 
1993).  The coincidence of nested resonances in the case of NGC 1530
further confirms the accuracy of the phase shift
method in determining the locations of the corotation resonances.

Also from Figure \ref{fg:fg1530res}, the {\it outer} intersection
of the lower $\Omega_p$ curve and the $\Omega-\kappa/2$ curve,
which signifies the outer-inner-Lindblad resonance (OILR) at radius
about 5 kpc, coincides with a depression spot on the phase shift curve
of Figure \ref{fg:fg1530} Left, indicating a region of low
angular momentum exchange between the bar density wave and the disk matter.
The same radius corresponds roughly to a region of low surface density  
as well on the image of NGC 1530 in Figure \ref{fg:fg1530}, right,
where the central oval and the straight bar section seemed to disjoin.

The nested resonances in the central regions of intermediate-to-early
type galaxies are found to be important to the channeling
of matter to fuel nuclear activity (Zhang et al. 1993).  Since
these nested resonances only form when the potential near the
central region of a galaxy is deep enough, the outer mass accretion,
which helps to transform a galaxy's morphology from late to early 
Hubble type (see \S 4.3 for further discussions), appear to serve 
as a preparatory stage for the fueling 
of the active nuclei.  The same mass accretion process across
a hierarchy of resonance features may also have placed an important
role in the formation of the tight black-hole-mass/bulge-mass 
correlation relation observed for many early type galaxies
(Gebhardt et al. 2000 and the references therein; 
Merritt \& Ferrarese 2001 and the references therein).

\subsection{Comparison with Corotation Radii Determined using the 
Tremaine-Weinberg (TW) Method}

As noted in \S 1, the derivation of galaxy pattern speeds and 
corotation radii
has historically been difficult.
The Tremaine and Weinberg (1984) method
is considered to be the most direct approach, which
uses the continuity equation
and off-nuclear spectra to derive pattern speeds, from which
a corotation radius may be inferred if the rotation curve is
known. Although widely applied, the method has numerous operational
difficulties that make it impractical in some circumstances:
(1) not easily applied to face-on galaxies;
(2) bar can't be end-on or broadside-on;
(3) off-nuclear spectra can be so time-consuming to obtain
that few galaxies have received the telescope time needed;
(4) the continuity equation must be satisfied for the
tracers being used, which is the reason the method has been 
applied mainly to SB0 galaxies, leaving spirals largely
unstudied; 
and (5) application is complicated and
model-dependent if there is more than one pattern present.

\begin{figure}[ht!]
\vspace{3.1in}
\includegraphics{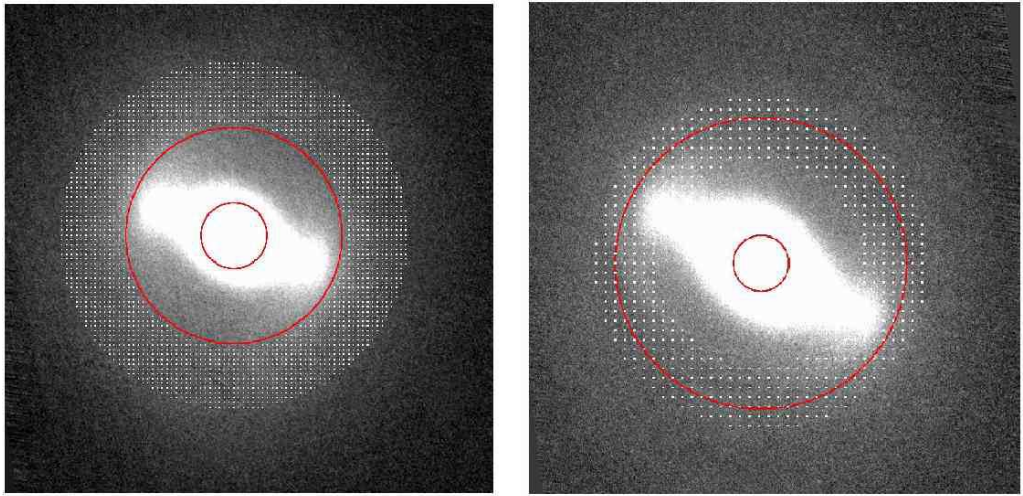}
\caption{Comparison of CR radii (circles) derived from
the phase shift method with CR bounds determined by
the Tremaine-Weinberg method (hatched regions) for
({\it Left}) NGC 936 (Merrifield \& Kuijken 1995) and ({\it Right})
NGC 4596 (Gerssen et al. 1999).
The phase shift results are based on $K_s$-band
(2.15$\mu$m) images (Laurikainen, Salo, and Buta
2005).}
\label{fg:fgTW}
\end{figure}

In Figure \ref{fg:fgTW} we plot a comparison of the corotation radii
calculated by the phase shift method (circles) with
the corotation bounds derived
by the TW method (hatched regions) for two early-type barred galaxies: 
the SB($\underline{\rm r}$s)0$^+$
galaxy NGC 936 (Merrifield \& Kuijken 1995) and the SB(rs)0/a galaxy 
NGC 4596 (Gerssen et al. 1999).  The TW method as applied to the 
above two galaxies was used to determine only
the outer pattern CR radius, while we find inner CR radii as well.
In each case, the phase shift method gives an outer CR radius in good 
agreement with the TW method.

\begin{figure}[ht!]
\plotone{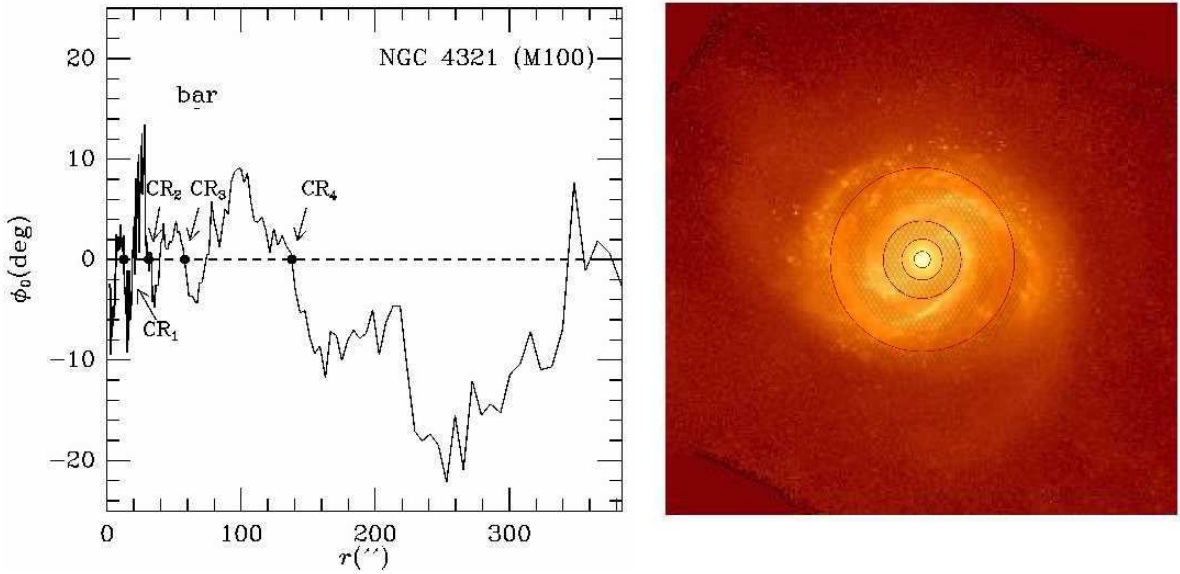}
\caption{{\it Left}: Phase shift plot for M100 (NGC 4321) 
based on a deprojected 3.6$\mu$m
SINGS image. {\it Right}: Comparison of the CR 
radii (4 circles) derived from
the phase shift method, with the CR bounds determined by
the Tremaine-Weinberg method (3 hatched regions) from Hernandez et al.
(2005) for M100.  The box size on the right is 12\rlap{.}$^{\prime}$8 
by 12\rlap{.}$^{\prime}$8.
These CR result comparisons are superposed on the 3.6$\mu$m Spitzer 
SINGS image 
of M100 (Kennicutt et al.  2003).}
\label{fg:fgTWM100}
\end{figure}

In Figure \ref{fg:fgTWM100} we plot a comparison of corotation radii
calculated by the phase shift method (circles) with
the corotation bounds derived
with the TW method (hatched regions, Hernandez et al. 2005))
for a late-type barred galaxy M100/NGC 4321.
The phase shift results are calculated using 
a Spitzer SINGS 3.6$\mu$m Legacy image (Kennicutt et al. 2003).
We find four well-resolved corotation radii for this galaxy, whereas 
Hernandez et al. (2005) previously derived
three corotation radii using a 2D velocity field and restricting
their analyses to the different regions assumed for different modes. 

The CR radius for the outer spiral pattern determined by both the 
Hernandez et al. analysis
(outer hatched region, 140"-160") and by the phase shift analysis (outer red
circle, 138") falls in the middle of the outer faint spiral arms, close to
where the inner, stronger spiral pattern appears to truncate.
For this CR the phase shift method and the TW method give results
which seem to be in good agreement.

The innermost corotation determined by the phase shift method (13"), 
which surrounds the nuclear bar-ring pattern, was not detected by the
Hernandez et al. analysis, possibly because of the lack of spatial
resolution for the data set used.
For the intermediate corotations, the phase shift method found two
(31" and 59"),
which sandwich the first corotation circle determined by Hernandez
et al. (32"-61"), shown as the innermost hatched region.
It is at this point open to question whether these two closely-spaced
intermediate crossings indicate two sets of 
quasi-steady modes with two different pattern speeds, or else
are caused by noise or the non-quasi-steady nature of the pattern.  
However, we found the middle
hatched region (80"-110")
as determined by Hernandez et al. to have no corresponding
feature in the phase shift calculation. This middle hatched
region was suggested by
Hernandez et al. to be the corotation of the bar of M100. 

\begin{figure}[ht!]
\vspace{5.1in}
\includegraphics{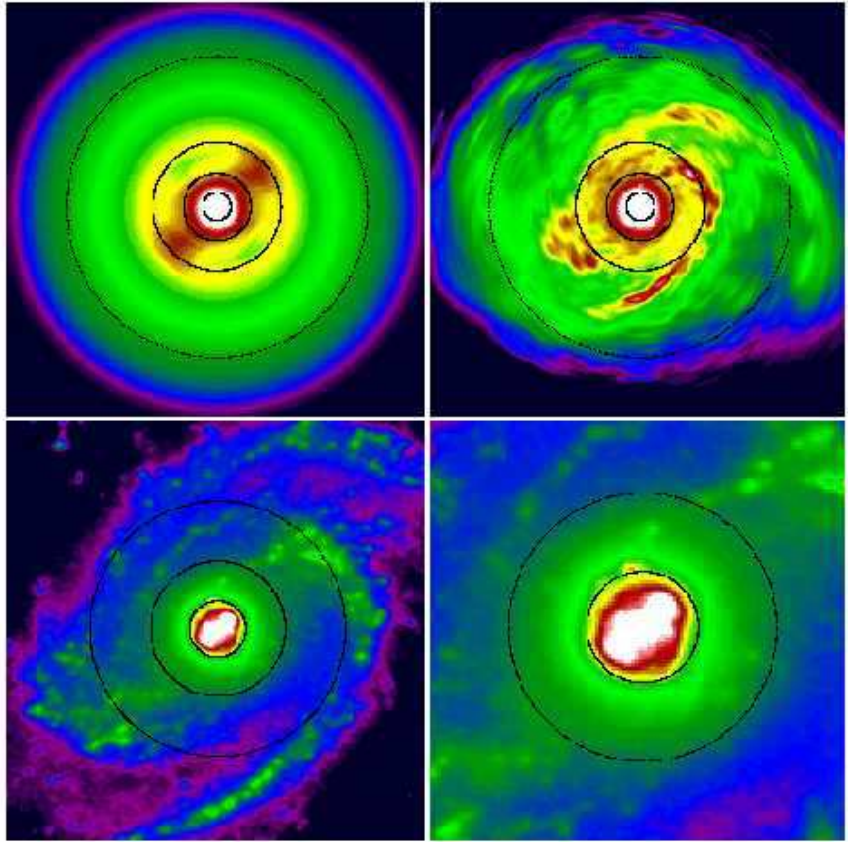}
\caption{{\it Top Left:} Bar-separated
SINGS image of the bright inner region of NGC 4321 superimposed
with the 4 corotation circles determined using the phase shift method.
The box size of the image is 6\rlap{.}$^{\prime}$4 by 
6\rlap{.}$^{\prime}$4.
{\it Top Right:} Spiral-separated
SINGS image of NGC 4321 of the same region as at left superimposed
with the 4 corotation circles determined using the phase shift method.
The box size of this image is also 6\rlap{.}$^{\prime}$4 by 
6\rlap{.}$^{\prime}$4.
{\it Bottom Left:} SINGS image (without bar-spiral separation) of 
NGC 4321 with a factor of 2 linear
zoom compared to the top panels (box size 3\rlap{.}$^{\prime}$2 by 
3\rlap{.}$^{\prime}$2), 
superimposed with the central 3 corotation
circles determined using the phase shift method.
{\it Bottom Right:} SINGS image (without bar-spiral
separation) of NGC 4321 with a factor of 4 linear
zoom compared to the top panels (box size 1\rlap{.}$^{\prime}$6 by 
1\rlap{.}$^{\prime}$6), 
superimposed with the central 2 corotation
circles determined using the phase shift method.}
\label{fg:fgzoom}
\end{figure}

To deduce the bar radius, we used the SINGS image to perform a
bar-spiral separation using the methods described by Buta et al.
(2005), the results of which are shown with our four CR circles
superposed in the top two panels of
Figure \ref{fg:fgzoom}. From the separated bar image, we estimate a bar 
radius
ranging from 65\rlap{.}$^{\prime\prime}$8, based on the isophote
of maximum ellipticity, to 68\rlap{.}$^{\prime\prime}$5, based on
the faintest-detectable bar isophote. The closest positive-to-negative
phase shift crossing to the bar radius is $r$(CR$_3$) = 
58\rlap{.}$^{\prime\prime}$1, which is slightly inside the bar ends.
We consider our corotation radius a more natural choice for the bar
than the Hernandez et al. determination, which is much further
out at 80$^{\prime\prime}$-110$^{\prime\prime}$ in the middle of the
bright inner spiral, whereas our CR$_3$ circle contains little of 
the main spiral pattern.  Our bar-size determination is only slightly 
longer than the value (63") obtained by Knappen et al. (1995).  Given
the intrinsic uncertainties in the bar-spiral-separation approach
these two determinations can be considered in perfect agreement.
Our CR$_3$ determination gives a value for the main bar feature
which is also close to the CR$_{bar}$ values determined by
Arsenault et al. (1988) and by Rand (1995).
 
The lower two frames of Figure
\ref{fg:fgzoom} present the original
image (without bar-spiral separation) zoomed-in by a linear factor
of 2 and 4, respectively, compared to the top two frames.  
From these two frames we can see more
clearly that the innermost corotation circle (CR$_1$) encloses
the strong secondary bar.  Between the next two CRs
(CR$_2$ and CR$_3$) there appear to be faint spiral structures.
The correspondence between the CR circles and morphological
patterns highlights the possibility that all 4 CRs could be real, although
we do not exclude the possibility that some of the structures
in the intermediate regions may be transient.

We mention also that Oey et al. (2003) used an HII region spatial isochrone
method to determine the corotation radii for M100.
They found for the outermost corotation a similar value 
(154") as Hernandez et al. using the TW method (140"-160"), 
as well as ours (138") using the phase shift
method, but they did not find in their data the intermediate corotation
location found by Hernandez et al. and others ($80"-110"$), 
which is also absent in our results.  The HII region isochrones
are expected to be fairly reliable for determining the corotation
crossings since the density-wave-triggered star-formation behavior 
has been well-confirmed from many previous studies.

In the process of calculation for this galaxy we have found that the large
amount of star formation in the M100 image does not affect the phase shift 
results.  If we remove many of the star-forming regions, we get almost 
the same CR locations.  This is apparently related to the fact that the 
potential as calculated through the Poisson integral is determined by the 
global distribution of matter, and is insensitive to the small-scale local 
variations in intensity.  

\subsection{Leading Arms in a Grand-Design Spiral Galaxy}

The identification of a leading spiral pattern is equivalent
to determining the sense of pattern rotation
(i.e., clockwise or counter-clockwise; which is not the 
same as the sense of winding, i.e.
S or Z shaped spirals).  Once we can be sure which
way the pattern is rotating, then the leading spiral
or bar identification can be done mostly by eye.
In that sense the leading spiral 
identification does not have to involve phase shift 
calculations at all, if there is some other means to 
identify the sense of pattern rotation.

\begin{figure}[ht]
\plotone{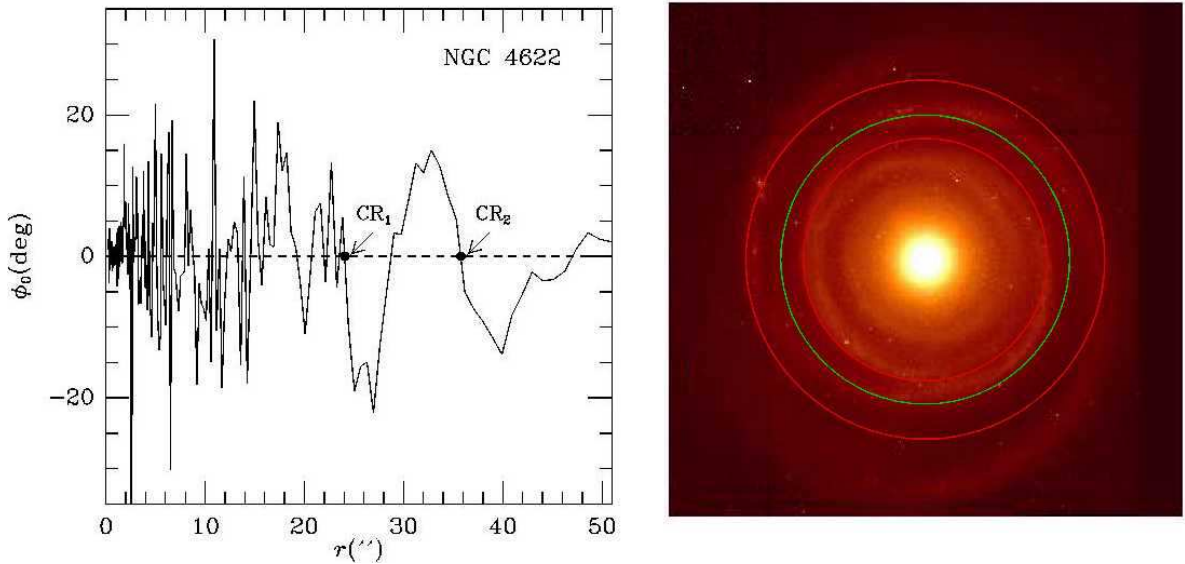}
\caption{
{\it Left:} Phase shift versus radius for the 
SA(r)a galaxy NGC 4622.  The phase shift
is plotted as ``-$\phi_0$'' because the galaxy 
is an S-sensed spiral where we have selected
the ``minus" plot rather than the usual plus plot. 
Thus, the sign convention is such
that the galactic rotation direction is clockwise.
Note also that this phase shift plot was made assuming a 2-armed
spiral, which is appropriate for the outer region.  
For the inner single-armed pattern, the actual phase shift
value should be doubled, although the zero-crossings would
remain the same.
{\it Right:} Log-scale $I$-band image of NGC 4622 with superimposed
corotation circles (red circles) of radii $24\rlap{.}^{\prime\prime}$1 
and 35\rlap{.}$^{\prime\prime}$8, and the green circle which
marks the major negative-to-positive crossing where the inner
trailing pattern and the outer leading pattern decouple. 
This image has not been corrected for the small inclination
(19$^{\circ}$) derived by Buta et al. (2003).}
\label{fg:fg4622}
\end{figure}

In the course of our study we have discovered an
unexpected application of the phase shift approach
for the confirmation of the leading nature of the
two outer spiral arms of the SA(r)a galaxy NGC 4622 (Figure
\ref{fg:fg4622}).  From the sense of winding of the outer arms, it would 
initially be placed in the ``S'' category, thus we should
have chosen to use the ``plus'' plot of phase shift versus
radius, if we assume the galaxy is rotating counter-clockwise
and thus the outer arms are trailing. With this choice
there are positive/negative crossings at $28\rlap{.}^{\prime\prime}7$ 
and $47\rlap{.}^{\prime\prime}$0.
The latter radius occurs in a no-data area so is ignored.
There is a strong negative/positive crossing 
in the outer arms at $35\rlap{.}^{\prime\prime}$8
if we assume this sense of galactic rotation. The 
$28\rlap{.}^{\prime\prime}$7  
circle lies between the inner and outer arms.

However, if we assume that the inner arm trails, the outer arms 
lead, and the galaxy 
rotates clockwise, as was recently demonstrated in Buta et al.
(2003), then there is a strong positive/negative crossing at 
$35\rlap{.}^{\prime\prime}$8 arcsec and a weaker
one at $24\rlap{.}^{\prime\prime}$1 arcsec. 
The former crossing is almost identical
to the corotation radius deduced in Buta et al. (2003)
by examining phase differences between $B$, $V$,
and $I$-band $m$=2 Fourier components.
So it appears that the clockwise rotation assumption leads
to a much more meaningful interpretation of the phase shift
data, and with results which also agree with the independent
analysis based on the dust/velocity field information used by
Buta et al. (2003).

If we did not have the dust/velocity field information,
independently we would not have been
able to deduce that the outer arms in NGC 4622 are leading.
Without the dust lane information and the corroboration
of the inner spiral pattern, it would be
a much weaker case to argue that the outer arms are leading just
from the phase shift results, because we really don't know
where corotation should be.  To make sense of the phase shift plot
we could equally have chosen the positive/negative crossing at 
$28\rlap{.}^{\prime\prime}$7
from the ``plus'' plot by assuming counter-clockwise galactic rotation. 

Since we had to use the ``minus'' plot for an S-shaped spiral,
and the corotation crossing determined from a positive-to-
negative transition agrees with other independent determinations, 
that confirms that we have identified the sense of rotation of the 
pattern correctly, since even for leading patterns, the angular
momentum density of the pattern is still negative
relative to the disk rotation inside corotation (i.e. the pattern
rotates slower than the stars inside corotation), and vice versa
outside, so a positive-to-negative phase shift change means
that the pattern deposits negative angular momentum to the disk
inside corotation, and vice versa outside, so
the wave can grow by an enhanced amplitude (Zhang 1998).

This does not imply that the leading spiral
under concern transfers angular momentum outwards.
It implies only that since inside corotation the wave rotates
slower than the stars, and outside corotation it is the other way
around, if we take angular momentum away from the stars,
more stars can join the wave motion (and thus enhance the wave
amplitude).  For a positive phase shift inside corotation, the
wave torques the stellar density backward, so the wave would be
self-enhancing for such a phase shift distribution. It says
nothing about the radial direction of angular momentum flow,
only the angular momentum {\em exchange} between
the wave and the disk stars in an annulus.  Of course,
the angular momentum lost by the stars at a particular annulus
inside corotation is gained by all the other stars which made
up the wave motion collectively, since gravity is long-range.
We do not need an actual ``flow'' to take away the angular momentum
from the stars in an annulus, though in the case of a true spiral
mode such a radial flow of angular momentum does happen 
(Zhang 1998) -- and it
might be happenning here too, only that the flow direction could 
either be inward or outward.
The important point here is not the radial angular momentum
flow direction, but rather the {\em radial gradient}
of such a flow that is revealed by the sign of the phase shift. If there
is such a flow, the gradient of the flow -- i.e. the difference
between the entering and exiting ports -- must be such that the
stars in an annulus inside corotation lose angular momentum, to
be consistent with a potential-density phase shift of positive sign.

NGC 4622 is a unique case among the galaxies we have analyzed
so far, in that the density wave patterns in this galaxy (both
the inner single trailing pattern, the intermediate ``arc'' just 
outside the CR$_1$, and 
the outer two leading arms) are likely to be entirely a result of 
tidal interactions.  In a recent paper by Byrd et al. (2006, submitted 
to AJ), these authors have found that both the inner and outer density 
wave features can be reproduced by the past plunging passage of a small 
companion 1/100 the mass of NGC 4622.  Furthermore, this model predicts 
that the halo of this galaxy is 10 times more massive than the disk -- 
which is likely to be one of the reasons that the disk intrinsic modes 
are not prominent.  Partly as a result of the massive halo, as well as 
the spherical bulge component in the central region, the phase shift 
results for this galaxy in the inner region are pure noise, and are
not coherent modal responses.  Our inner crossing (24") 
is close to the value (21.5") determined by other independent methods as 
outlined in Buta et al. (2003) and Byrd et al. (2006), though not 
identical, perhaps reflecting the fact that the density waves in this 
galaxy are not intrinsic modes of the disk and had not reached 
quasi-steady state.

In general, one should always start
by assuming that the outer pattern is trailing (which had
been found to be the case for all spontaneously formed
N-body spiral modes),  and then see
if the phase shift plot and its predicted corotation radii
make sense when overlaid on the image.  Of course, in cases
like NGC 4622 where the inner and outer patterns have
oppositely winding senses, it is obvious that one set of the
patterns must be leading.  The question then becomes to
determine which one.  This still assumes that the inner and
outer pattern rotate in the same direction.  There exist
cases where even the rotation direction of the inner and
outer part of the disk stars (and possibly also density
waves) are different.  So in general we will need 
supplementary information, in addition to phase shift
calculations, to determine the sense of galactic rotation, and
the sense of leading and trailing density wave patterns.

An additional point to note is that for this galaxy the spiral 
patterns present are likely to be only transient density {\em waves},
instead of density wave {\em modes}.  That the phase shift method
also works in this case is fortuitous for us, and
in other cases of transient waves it should be expected that
the method may not give correct corotation radii if the pattern
is going through rapid morphological changes.

\subsection{Trend of Bar/Corotation Sizes versus Morphological Type}

An important issue in density wave
pattern speed studies is the ratio of the
corotation radius, $r_{CR}$, to the bar radius, $r_{bar}$. If this 
ratio is less than 1.4, a bar is considered ``fast", while if it
is larger than 1.4, a bar is considered ``slow" (Debattista \& Sellwood 2000).
Here the words fast or slow is in comparing the bar pattern speeds to
the local circular speeds at the location of the bar ends.

\begin{figure}[ht!]
\bigskip
\vspace{2.5in}
\centerline{
\includegraphics{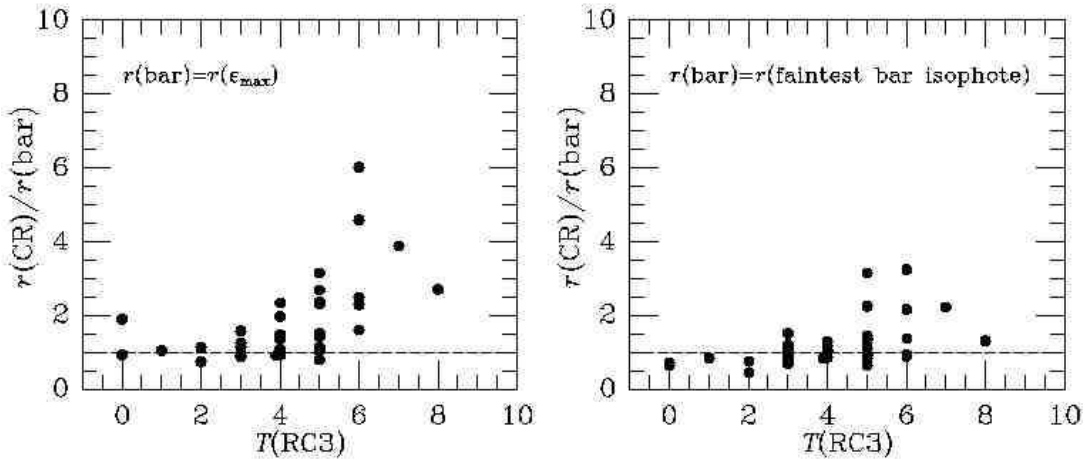}
}
\caption{Ratio of phase-shift-derived CR radius to bar radius for 
36 strongly-barred galaxies in the OSUBGS, based on bars that have been 
separated from their spirals (Buta et al. 2005). {\it Left:}
In this plot the bar radius was derived from the isophote 
having maximum ellipticity. {\it Right:} In this plot, the
bar radius was derived from the faintest distinct isophote. 
For this definition, over the entire sample,
$<r(CR)/r(bar)>$ =1.26 +/- 0.11.  
In these plots, T(RC3) is the numerically-coded RC3 Hubble 
type, where higher T represents later type, 
i.e. T=0 is S0/a, T=1 is Sa, T=2 is Sab, etc.}
\label{fg:fgsample}
\end{figure}

We have examined the
ratio of r(CR)/r(bar) in the OSUBGS sample using images where
the bar had been separated from the spiral using a systematic approach
(Buta et al. 2005).
The result is given in Figure \ref{fg:fgsample}.
Two definitions of the bar radius are used: the radius of
maximum ellipticity (left panel), and the radius of the lowest 
intensity
bar contour (right panel). The true bar length
should be bracketed by these two approaches.
The plots include only those objects
for which the relative gravitational bar torque strength
$Q_b$ $>$ 0.25 (Buta et al. 2005), 
to minimize some of the
confusion associated with weaker bars.

From Figure \ref{fg:fgsample} we see that in general the
bar length increases as the Hubble type of a galaxy changes
from late to early, converging towards
$r(bar) \approx r(CR)$ for the earliest morphological types,
a result also previously found by other investigators
(Elmegreen \& Elmegreen 1985; Rautiainen et al. 2005 and references therein).

\begin{figure}[ht!]
\plotone{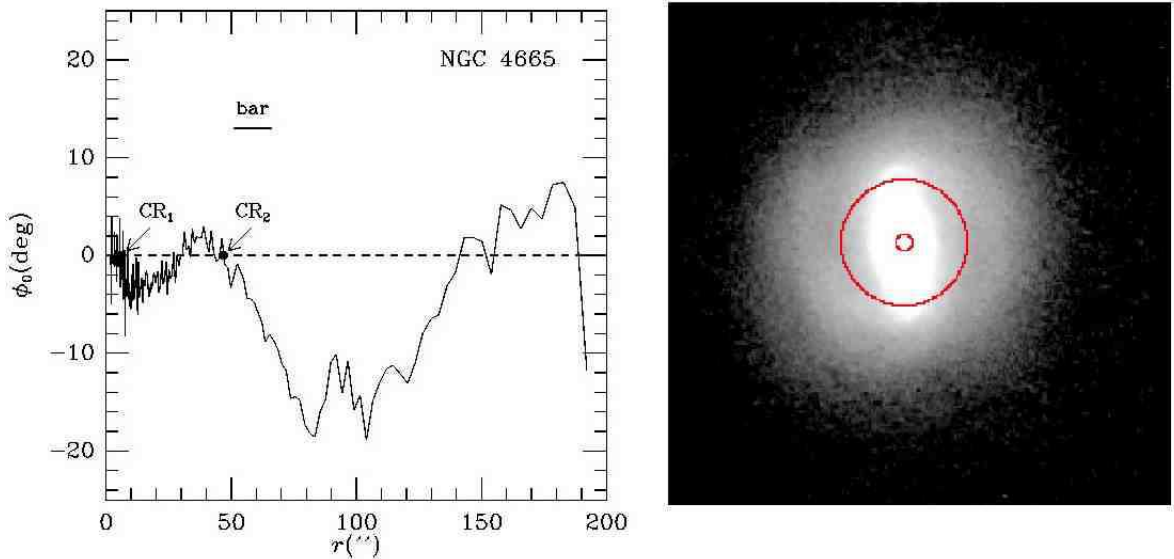}
\caption{
{\it Left:} Phase shift versus radius for the SB(s)0$^+$ galaxy NGC 
4665,
indicating one major positive-to-negative crossing, as well as
an additional inner crossing.
{\it Right:} Deprojected log-scale $H$-band image of NGC 4665
with superimposed CR circles (in red).}
\label{fg:fg4665}
\end{figure}

Among the galaxies calculated for the OSUBGS samples,
we have encountered a few cases where the bar appears to
extend beyond its CR circle (the inner bar of NGC 4321 which
we have encountered before is a mild case of this category), 
a phenomenon which seems to
challenge the so-far-well-accepted wisdom of the extent
of bars in galaxies (Contopoulos 1980).  Figure \ref{fg:fg4665}
gives one such example: the SB(s)0$^+$ galaxy NGC 4665, shown
with an $H$-band image from the OSUBGS sample.
From this figure it is obvious that this well-defined
bar pattern extends beyond its own corotation radius.

\begin{figure}[ht!]
\plotone{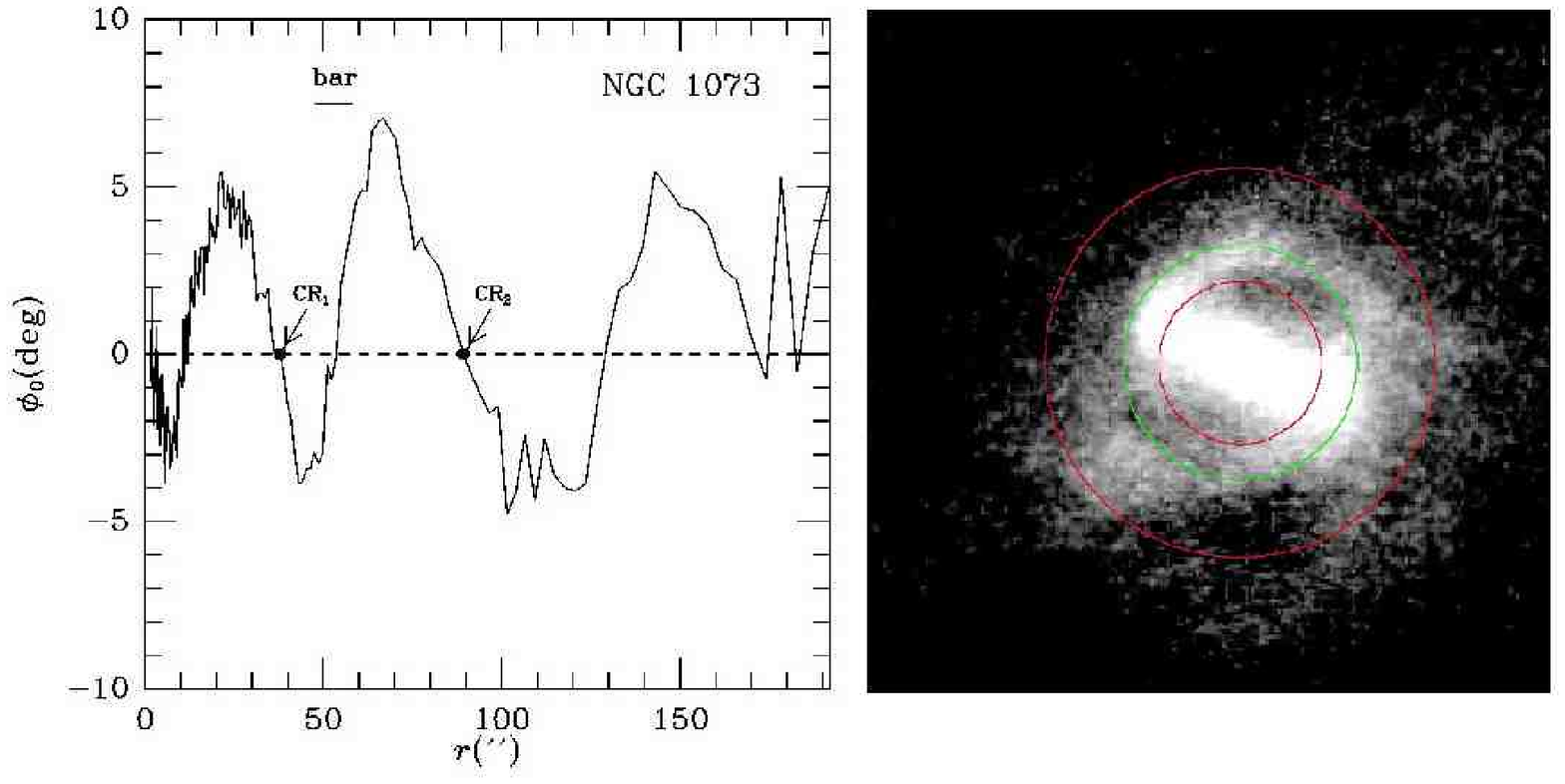}
\caption{ {\it Left:} Phase shift 
versus radius for the SB(rs)cd galaxy NGC 1073. Two CR radii are 
indicated,
neither of which coincides with the bar ends (short horizontal line). 
Instead, these ends appear
to be close to a major negative to positive crossing.
{\it Right:} Deprojected log-scale $H$-band image of NGC 1073 with 
superimposed
CR circles (in red). The green circle shows the major negative to 
positive
crossing.}
\label{fg:fg1073}
\end{figure}

An even more extended bar is shown in Figure \ref{fg:fg1073}
for OSUBGS galaxy NGC 1073.  Here the bar appears to
extend all the way to the
negative-to-positive crossing on the phase shift plot,
a location likely to be coinciding with the OLR of the inner bar.
Other cases which appear to fall into this category
include OSUBGS galaxies
NGC 3513, NGC 7741, NGC 4902, NGC 4781, and NGC 4579.
We will comment more on the possible dynamical mechanisms enabling
the longer bar extension in \S 4.1.

Note that for both the cases of NGC 4665 and
NGC 1073, there is a slight ambiguity
of how one defines the ``bar''.  For NGC 4665, the bar is slightly twisted
in its outer region to appear somewhat spiral-like.
For NGC 1073, on the other hand, at the nominal corotation circle
there is a slight ``pinch'' in the width of the pattern, with the inner bar
within the corotation circle having nearly ellipsoidal
isophotes, whereas the morphology of the outer extension of the bar
is more elongated, and has reduced surface density
compared to the portion inside $r_{CR}$.  Despite this ambuguity
in classification, one thing is certain that
the entire ``bar'' structure, both inside and
outside corotation, should be rotating with a single pattern speed.

We wrap up the current section by summarizing in Table 1
the numerical values of the bar lengths and corotation radii of the 
few galaxies we had analyzed quantitatively in this work.
The average of the ratios of CR radius to bar radius 
for this small sample of 7 barred galaxies (excluding NGC 4622) are:
$<$r(CR)/r(bar)$>$=1.01$\pm$0.15 if we use the radius of maximum
ellipticity for the bar, and  
$<$r(CR)/r(bar)$>$=0.87$\pm$0.14
if we use the ``faintest isophote'' for the bar. 
The detailed values for other galaxies in the list of
36 strongly-barred galaxies, as well as for other galaxies in the 
OSUBGS, will be given in the sequel paper (Buta \& Zhang 2007).

\begin{deluxetable}{lrccccc}
\tablewidth{0pc}
\tablecaption{Summary of Corotation and Bar Radii\tablenotemark{a}}
\tablehead{
\colhead{Galaxy} &
\colhead{$r$(CR$_1$)} &
\colhead{$r$(CR$_2$)} &
\colhead{$r$(CR$_3$)} &
\colhead{$r$(CR$_4$)} &
\colhead{$r$(bar)} &
\colhead{$r$(CR$_{bar}$)/$r$(bar)}  \\
\colhead{} &
\colhead{(arcsec)} &
\colhead{(arcsec)} &
\colhead{(arcsec)} &
\colhead{(arcsec)} &
\colhead{(arcsec)} &
\colhead{}  \\
\colhead{1} &
\colhead{2} &
\colhead{3} &
\colhead{4} &
\colhead{5} &
\colhead{6} &
\colhead{7}
}
\startdata
NGC  936 & 15.9 &  {\bf 52.0} & ...... & ...... &  47.8-55.0  & 
1.09-0.95 \\
NGC 1073 & {\bf 38.0} &  89.3 & ...... & ...... &  48.1-58.0  & 
0.79-0.66 \\
NGC 1530 &  5.8 &  {\bf 67.1} & ...... & ...... &  55.1-66.1  & 
1.22-1.02 \\
NGC 4314 &  9.0 &  {\bf 76.9} & ...... & ...... &  73.5-81.8  & 
1.04-0.94 \\
NGC 4321 & 12.5 &  31.0 &  {\bf 58.1}  & 138.0  &  65.8-68.5  & 
0.88-0.85  \\
NGC 4596 & 13.3 &  {\bf 69.9} & ...... & ...... &  63.0-73.6  & 
1.11-0.95 \\
NGC 4622\tablenotemark{b} & 24.1 &  35.8 & ...... & ...... &  .........  
& ......... \\
NGC 4665 &  5.4 &  {\bf 47.2} & ...... & ...... &  51.0-66.0  & 
0.92-0.71 \\
NGC 5247 & 13.9 &  86.9 & ...... & ...... &  .........  & ......... \\
\enddata
\tablenotetext{a}{Col. 1: galaxy name; cols. 2-5: radii of positive to 
negative
phase shift crossings, taken to be CR radii. Numbers in boldface are 
taken to be bar CR radii. All of the galaxies have inner CRs associated 
with 
inner spirals, secondary bars, or ovals, and NGC 4321 shows 4 prominent
CR crossings corresponding to well-recognized morphological
features. Several galaxies show additional
crossings in the very centers of the nuclear region which were barely 
resolved by the phase shift calculations (NGC 4314, NGC 5247, NGC 1530, 
NGC 4321, and NGC 1073), the values for these crossings were not 
explicitly 
given due to the insufficient resolution at these galactic central 
regions; 
col. (6): estimates of the bar radius based on two methods:
fitted radius of the ellipse of maximum ellipticity (smaller value), 
and radius
of the faintest-detectable isophote (larger value), based on an image 
where
the bar has been separated from its associated spirals or rings. 
Several
galaxies have certain degree of asymmetry between  the two sides of the 
bar,
such as NGC 1073; col. (7): ratio of
CR to bar radius based on the range in col. 6.}
\tablenotetext{b}{The CR radii are based on a graph of $-\phi_o$ versus 
radius (see text).}
\end{deluxetable}

\section{ANALYSIS AND DISCUSSION}

\subsection{Extent of the Spiral and Bar Patterns}

Although during the early stage of the density wave development
it had been suggested (Lin 1970) 
that the spiral pattern should terminate
at corotation, this conclusion was subsequently
superceded by improved theoretical analyses
(Mark 1976; Lin \& Lau 1979; Toomre 1981), observational
results (Canzian 1998 and the references therein), and
N-body simulations (Donner \& Thomasson 1994; Zhang 1996),
which all showed that spiral density waves can extend to
as far as the OLR of the pattern, though a fraction
of the observed spiral galaxies indeed displayed
evidence of a significant reduction of surface brightness
beyond the corotation radii (Zhang et al. 1993; Oey et al. 2003).

For barred galaxies, a similar conclusion (that the pattern
terminates at its own corotation) had been reached by
Contopoulos (1980), based on orbit analysis under an applied
potential to construct
``self-consistent'' models of barred galaxies.  It was found
that the phases of the periodic orbits thus obtained
usually change rapidly when approaching
the corotation radius (the phase change starts to be significant
around the 4:1 resonance which lies within corotation), 
making such self-consistent model constructions
difficult.  This was the underlying reason for Contopoulos (1980)'s
conclusion that a bar should terminate at corotation. 

Since it is by now well-established that a spiral pattern
can extend beyond its corotation radius, and since
bars are density wave patterns just like the spirals
apart from the fact that they reside in the central regions 
of galaxies, one naturally suspects that a bar might be 
able to extend beyond its own corotation radius as well.
This possibility in fact was hinted already in Contopoulos
(1980), where the orbital behavior of stars near
the inner-Lindbald-Resonances (ILRs) 
of a galaxy was analyzed in addition to that
near the corotation.  In particular, for galaxies
which possess two ILRs, even though normally the inner bar ends
within the IILR, in certain cases ``the inner
bar can extend up to the outer ILR if its amplitude
drops sufficiently fast" (Contopoulos 1980).

For nested resonances, it often happens that the corotation
of the inner secondary pattern is close to the location of
the IILR of the outer pattern (as we have seen for the
case of NGC 1530), and the OLR of the secondary pattern
is close to the OILR of the outer pattern (for
the case of NGC 1530 though, the OLR of the inner pattern
resides within the OILR of the outer pattern). Therefore,
Contopoulos's conclusion that the
secondary pattern can sometimes fill in the region
between two ILRs implicitly admits that the inner
secondary bar can extend beyond its own corotation,
and reach all the way to its own OLR, and perhaps
even beyond (if the OLR of the inner pattern lies within the OILR
of the outer pattern).  

The evidence that bars can extend beyond its corotation had in fact already
showed up in the earlier self-consistent N-body simulations.
Sparke and Sellwood (1987) had produced one of the earliest
self-sustained 2D N-body bar modes which settled onto quasi-steady
state after first emerging as a spiral pattern.  During
much of the simulation run after the mode had reached nonlinear saturation
amplitude, the extent of the bar in fact
is longer than the corotation radius at the corresponding time step.  
For example, from their Figure 3(a), we can see that the inner bar mode 
has contours extending to about $1.7$ in radius at time step 40, 
whereas the corotation radius for the same time
step is of size $1$. In a later stage (step 103) of the simulation 
(their Figure 3b) the corotation grows to $1.3$ (due to the slowing 
down of the bar pattern), and
the bar on the other hand had reduced in length so it is indeed 
enclosed within the corotation radius at this late stage
of the simulation.  Note that even though the morphology of the bar
and spiral pattern in the Sparke \& Sellwood (1987) simulation is
evolving continuously, this evolution is the evolution
of the quasi-steady modes as a result of the evolution of the underlying 
basic state distribution due to the wave-induced
angular momentum transport and exchange.  The wave modes themselves
had achieved quasi-stady state throughout most of the simulation run,
as reflected in the well-defined pattern speeds that delineate
precisely the extents of the inner and outer modes (see their Figure 
3). That the Sparke \& Sellwood (1987) simulation had the longest
bar length (compared to the size of the modal corotation radius)
at the intermediate stage of the simulation is also consistent with
our finding that the smallest r(CR)/r(bar) ratios are observed for
intermediate Hubble type galaxies.

We will further argue in \S 4.2 that the fact that bars and spirals can 
extend beyond their own corotation radii is partly due to the fact 
that when self-consistent patterns and collective effects are 
taken into account, many of the conclusions reached through passive
orbit calculations are no longer valid.
As a matter of fact, there will in general not be a clear
dividing line between bars and spirals (apart from the fact that bars
tend to occur in the inner regions of galaxies and spirals the
outer regions): most realistic bars in physical galaxies
(and this includes the more than 100 galaxies we have examined
so far in the OSUBGS sample) all
possess a certain degree of skewness\footnote{Certain very early type 
galaxies
may contain very limited amount of sknewness (and thus a very small
resulting phase shift), such as many S0 and disky elliptical galaxies.
These galaxies can be considered old or even close to being``dead''
in the evolutionary sense, in that the phase-shift-induced secular
evolution had practically halted.}, which enables the presence of the
phase shift, and the associated collective dissipation effect which leads
to the secular evolution of the basic state (see \S 4.3).
In cases of bar-driven spirals it might even be hard to tell where the 
bar ends and the spiral starts (such as the case of NGC 4321), especially 
during the intermediate stage of the evolution when the spiral is 
being transformed into a bar-driven spiral.  For other galaxies
(such as NGC 1073, NGC 4665, the inner mode of NGC 1530,
as well as the simulated galaxy disk in Sparke and Sellwood [1987]),
the scenario of bar-extending-beyond corotation is achieved through
the decoupling of the inner and outer nested modes.

\subsection{Orbits as ``Building Blocks'' of Large-Scale
Density Wave Patterns}

Stellar trajectories in a self-consistent spiral pattern can
show drastically different behavior than periodic or quasi-periodic
orbits calculated in an applied spiral potential.
While both groups become chaotic at large spiral
amplitudes, the former also exhibit secular mean-radius change, 
which is not displayed in the latter.  Such a difference in the 
behavior is due to the constraint of the potential-density
phase shift applied by the self-consistent spiral pattern on 
the individual stars' trajectories, which causes the spiral wave to 
steepen into large-scale (collisionless) spiral shocks.
This further leads to the irreversible conversion of stellar orbital 
motion energy into random motion energy, as well as into the energy 
of the wave, through a local gravitational instability of the 
streaming disk material at the spiral wave crest (Zhang 1996,1998).
We now present a quantitative example to illustrate the difference
in behavior between a passive orbit calculated in a rigidly-rotating,
applied spiral/bar potential, and a stellar trajectory obtained
from a self-consistent N-body calculation.

\begin{figure}
\vspace{7.in}
\centerline{
\includegraphics{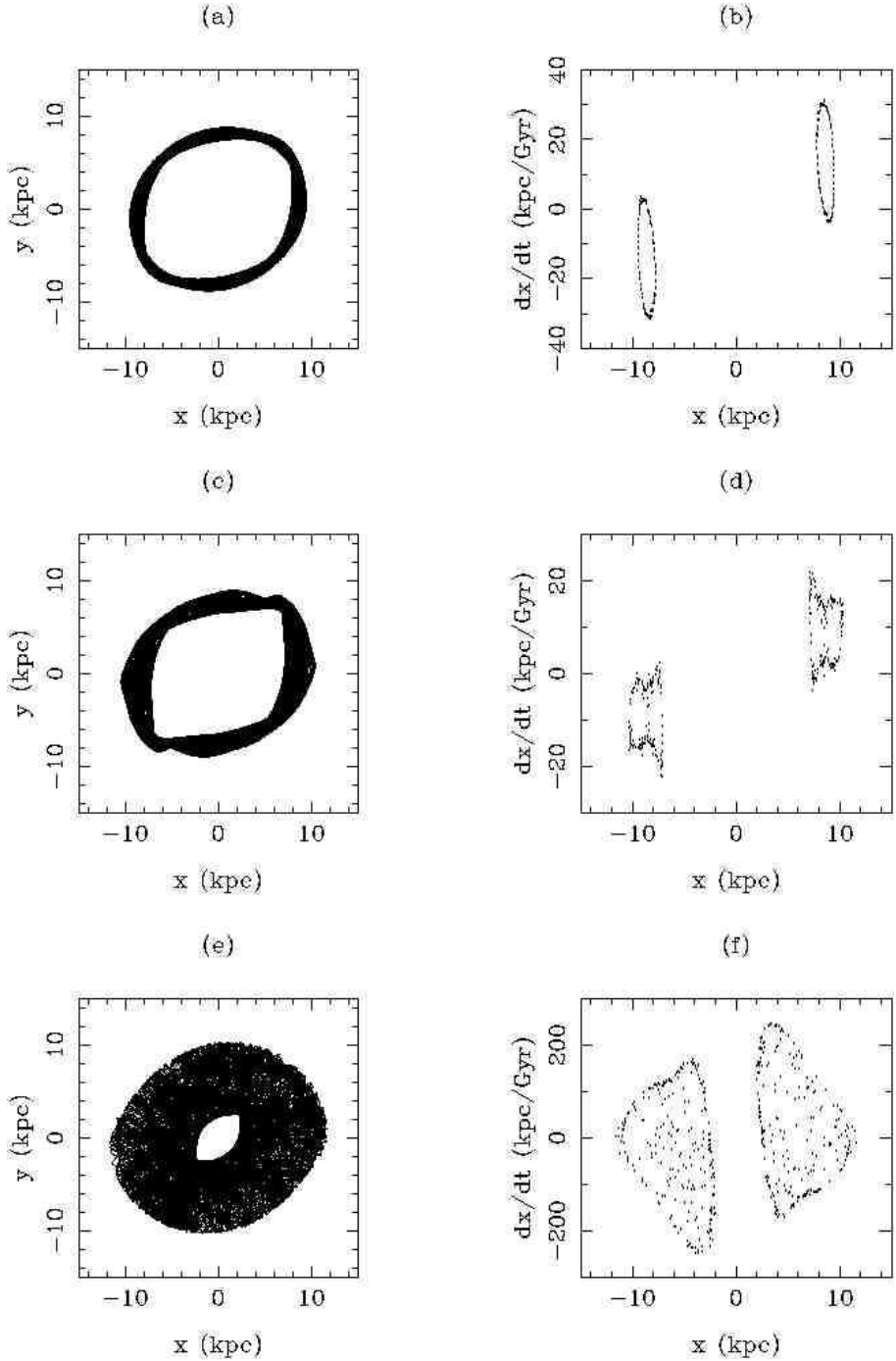}
}
\caption{Orbit solution, plotted in the pattern-stationary
frame, and surface of section
for spiral forcing strengths
of 10\% (a and b), 20\%
(c and d), and 30\% (e and f), respectively.
Parameters common to these three cases are: the pitch
angle of the spiral $i$= $20^{o}$, $\Omega_p = 13.5$ km s$^{-1}$ 
kpc$^{-1}$,
$r(t=0) = 8.5$ kpc, $v_c = 220$ km s$^{-1}$, and
total integration time t = $80$ Gyr.
The test particle was initially on a circular orbit.}
\label{fg:fgchaos1}
\end{figure}

In Figure \ref{fg:fgchaos1}, we plot the nonlinear orbit solution, 
together with the corresponding surface of section,
for spiral forcing of strength $10\%$, $20\%$, and
$30\%$, respectively, under an axisymmetric mean
potential similar to that in the solar neighborhood.
The equations used for calculating the orbits
are the standard nonlinear coupled second-order
ordinary differential equations in the corotating
frame of the spiral (Binney \& Tremaine 1987
equations [3-107a], [93-107b]), i.e.

\begin{equation}
\ddot{r} - r {\dot{\phi}}^2 = - {{\partial \Phi} \over {\partial r}}
+ 2 r \dot{\phi} \Omega_p + \Omega_p^2 r
,
\end{equation}
\begin{equation}
r \ddot{\phi} + 2
\dot{r} \dot{\phi} = - {{1} \over {r}}
{{\partial \Phi} \over {\partial \phi}}
- 2  \dot{r} \Omega_p
,
\end{equation}
and with the forcing potential being of the form
\begin{equation}
\Phi = {{v_c^2} \ln r} + A \cdot \cos \left(2 \phi +
{{2 \ln r} \over {\tan i}} \right)
,
\end{equation}
where $v_c$ is the circular speed, $\Omega_p$ is
the pattern speed and $i$ is the pitch
angle of the spiral. In all the calculations performed for
Figure \ref{fg:fgchaos1}, we have used
$r= 8.5$ kpc, $v_c = 220$ km s$^{-1}$, $\Omega_p = 13.5$
km s$^{-1}$ kpc$^{-1}$,
and $i = 20^o $. The fractional forcing strength $f$ is defined as
$f  \equiv 2 A/(\tan i \cdot v_c^2)$.

We observe from Figure \ref{fg:fgchaos1} that for all the orbits
calculated, there is no secular change in the mean orbital radius,
which is a result consistent with the constancy of the Jacobi integral
in a rigidly-rotating spiral potential (Binney \& Tremaine 1987, equation 
[3-88]).  Furthermore, we notice that orbits are only regular at small
spiral forcing amplitude. For forcing amplitudes which exceed 20\%,
the orbits gradually become chaotic, and their traces on the surface 
of section begin to cover an entire two-dimensional area.

It is obvious that a coherent spiral density distribution
can not be obtained by the straightforward superposition of orbits of
the type given by Figure \ref{fg:fgchaos1}(e).  However, 20-30\%
is a realistic forcing level in the observed spiral and barred
galaxies.  Therefore at least in the 
case of (but not limited to) large amplitude spirals, orbits 
can no longer be considered as the ``building-blocks'' of the global 
spiral pattern.  Certain collective effects have to be present in order to
``collapse the chaos'' inherent in the stellar trajectories,
to arrive at a coherent spiral density response.
Contopoulos and collaborators had in later years also
included more of the role of chaotic orbits in constructing
galaxy models (Contopoulos \& Voglis 1996 and 
references therein; Voglis, Stavropoulos \& Kalapotharakos 2006), 
though the full collective effects are still
missing in these passive calculations\footnote{The chaos present in the passive
orbit response is in general of a different nature than the chaotic
behavior observed in collective systems, i.e., it lacks the hidden
collective correlations among the N-particles' trajectories in
collective systems, yet displays more coherent patterns and correlation
with respect to the mean forcing potential.}.

\begin{figure}
\vspace{7.in}
\centerline{
\includegraphics{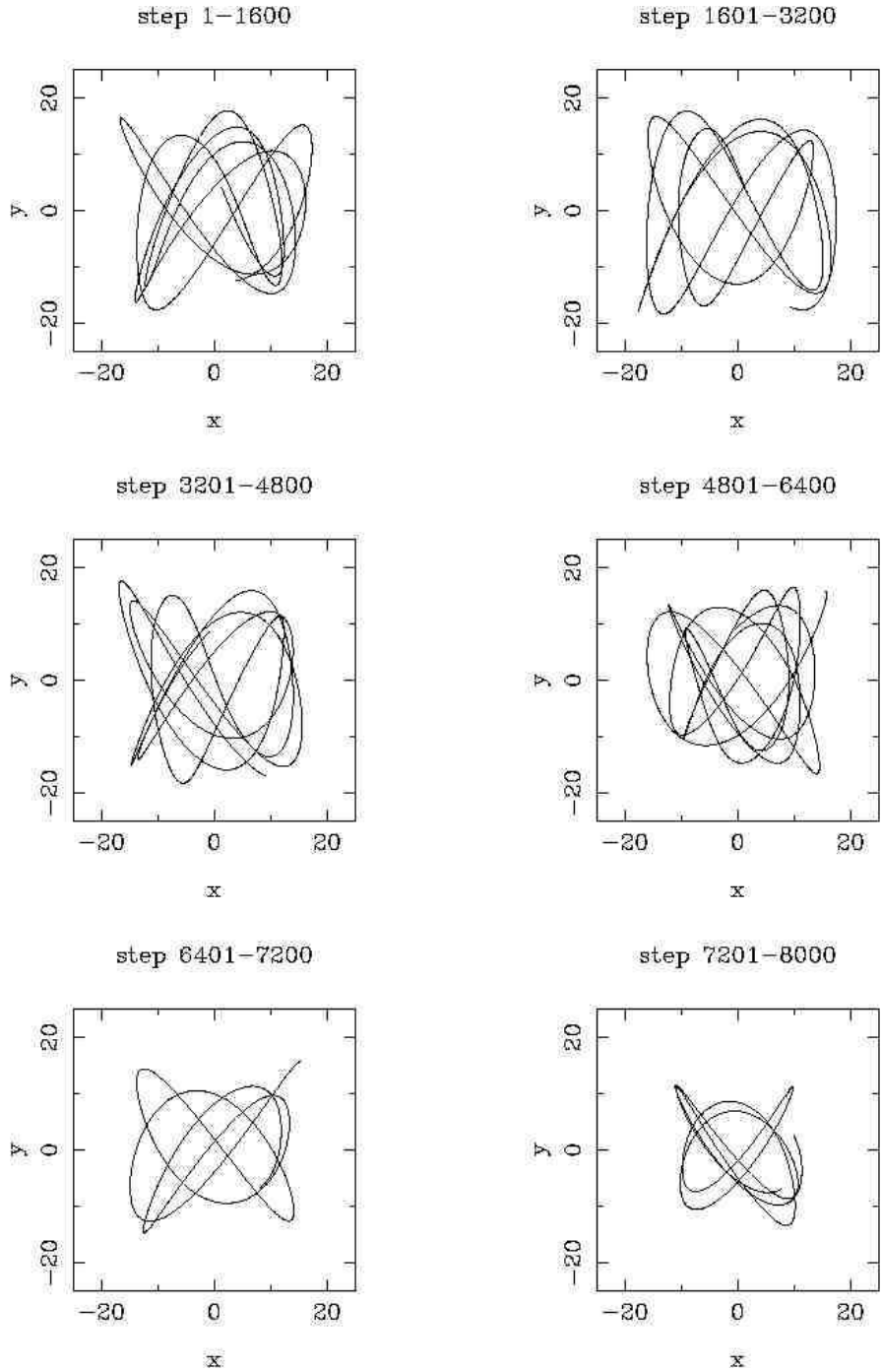}
}
\caption{Time evolution of the trajectory of a typical star inside 
corotation, for a spontaneously formed N-body spiral
mode simulated in Zhang (1996).
The trajectory sections are plotted with respect to
the pattern-stationary (corotating) frame.
The forcing amplitude is around 15\%.
}
\label{fg:fgchaos2}
\end{figure}

N-body simulations of spontaneously-formed spiral modes revealed that 
typical particle trajectories in such finite-amplitude, self-sustained 
spiral patterns are again chaotic in the corotating frame of the pattern; 
however, a new phenomenon not present in the passive orbit calculations
is now also evident.  A typical trajectory of a particle in the 
simulation of Zhang (1996) in the corotating frame
of a quasi-stationary spiral pattern,
for a radial location inside corotation, is plotted in Figure 
\ref{fg:fgchaos2}.  This simulation run obtained spontaneously
formed spiral mode of equilibrium amplitude around 15\%.
It is clear that this particle trajectory covers a two-dimensional area
in the reference frame corotating with the density wave pattern,
signaling the presence of chaos (without chaos, for a 15\% forcing
the orbit should look like the tubes in Figure \ref{fg:fgchaos1}, between the
top two frames in the left column).  The invasion of 
chaos is thus found to occur at a much smaller spiral forcing amplitude
for the spontaneously formed N-body spirals than for
the non-self-consistent case, a result consistent with the theoretical
finding that collective relaxation induced by large-scale global
instability patterns generally dominates the relaxation behavior of
stellar trajectories in galaxies (Weinberg 1993; Pfenniger 1986; Zhang 
1996). 

Another new phenomenon evident from Figure \ref{fg:fgchaos2} is
that there is now a secular decay of the mean orbital radius for stars
inside corotation. A corresponding secular increase of the mean orbital 
radius for stars outside corotation is also found (Zhang 1996).
Both of these secular changes in mean trajectory sizes are
absent from the non-self-consistent forced orbit calculations.

The presence of stochasticity in an individual star's orbit is an 
integral element in a self-organized and self-sustained spiral pattern.  
The stochasticity derives from the exponential sensitivity of
the stellar trajectory to the small perturbations in initial/boundary
conditions, with the sensitivity itself a result of the presence
of global instabilities which creates an effective ``collisional
relaxation'' condition (Zhang 1996).  The large-scale coherence of a 
spontaneously-formed spiral pattern is nonetheless not sensitive 
to the presence of the chaotic behavior in an individual star's
trajectory.  This is because the spontaneous emergence, and the
subsequent maintenance and secular evolution of these coherent
patterns is a natural sequence which corresponds to a large phase space volume 
in all the possible initial conditions that a proto-galactic-disk could
choose from.  The large increase in coarse-grained
entropy when a disk develops a spiral mode means that 
the pattern itself is an {\em attractor} to the evolution of these
particles' collective motions.

Therefore in spiral disks we witness three levels of organization:
deterministic in the governing equations; chaos in the individual 
star's trajectory; and deterministic again in the large scale patterns 
formed in such a many-body system. Collective dissipation is what 
made possible the existence of large-scale coherence in a sea of 
underlying chaotic orbits.  Each time at the crossing of the spiral
or bar crest, the orbital parameters of the individual stellar
trajectories are re-organized to support
the wave pattern, and they are completely de-correlated from their
parameters in the previous cycle of arm-crossing, due to the
presence of gravitational instability at the arms.
The presence of global resonances (such as ILRs and CRs)
in such self-consistent patterns will no
longer act on individual orbits, but rather on
the collective motions of all the particles 
constituting the density wave\footnote{The
fact that an individual star's resonance condition is totally destroyed
in a disk galaxy containing a global instability can be understood
as due to the decorrelation of the phases of the trajectory each time
it experiences a gravitational-instability-induced scattering/collision
event at the spiral arm crest.  Like the example of a child's swing, both
the exact frequency condition and the exact phase condition need to
be satisfied for the swing to gain in amplitude.  In the case of
individual trajectories in galaxies, it is the phase coherence
condition that is being destroyed by the density-wave-induced 
gravitational instability.}. In a sense, the collection of stars
behave rather like molecules in a fluid, and the collisional shock
in the spiral arms and bars create conditions of particle interaction
akin to true collisions in a true fluid, albeit here not in the sense 
of two-body collision, but rather the scatterings of particles in 
the collective instability at the density wave crest.

So far, all the existing orbital resonance calculations 
and galaxy model constructions were done under an
applied potential, and not the kind of self-consistent potential in
the many-particle systems.  Even for several supposedly
self-consistent calculations using N-body simulations, a closer examination
of the ways some of the final analyses were conducted revealed
that it was only the forced/passive orbital behavior that had
been uncovered in these analyses.  For example, Athanassoula
(2003) studied the resonance exchanges between bars
and active halos.  Even though she had started out with a self-consistent
three-dimensional N-body simulation, when the time comes to 
calculate the resonant exchanges between the bar-stars and halo-stars,
she had to halt the N-body simulation and revert to using
a {\em rigid}, {\em uniformly rotating}, and {\em enforced}
bar potential which best approximates the bar
potential she had obtained at the end of the N-body simulation,
so as to obtain the sought-after orbital resonance
exchange with the halo. It is not surprising that such a
practice should be needed: for resonance conditions exist for individual
orbits only under an {\em applied} and {\em rigidly rotating}
potential, and not for a self-consistent and (microscopically)
time-fluctuating potential from the 
actual N-body simulations where collective effects completely
wipe out the individual orbit's resonance behavior.  Likewise,
other investigators who had conducted periodic-
and quasi-periodic-orbit-searches in self-consistent N-body
models had also used either {\em the N-body potential
at one instant of the simulation} (Pfenniger \& Friedli 1991)
or a {\em time-averaged potential} (Sparke \& Sellwood 1987)
to search for periodic orbit families, which effectively
are equivalent to using a forced, rigidly-rotating and
non-self-consistent potential over the orbital period under study.
It is no wonder that the secular mean trajectory changes we have
found in the true, fully-self-consistent N-body simulations were
never found in those orbit searches which used the N-body potential
at a fixed instant to approximate the potential over the entire orbital 
period.

The true collective behavior is encoded in, and is enabled by,
the spatial and temporal correlations of the N-particles'
positions and velocities.  While the phase correlation of a single star's 
trajectory with the mean forcing potential has been destroyed
by the presence of collective instabilities\footnote{Note
that this appears to have violated the conclusion of the celebrated
KAM theorem (after Komogorov, Arnold and Moser), which prescribes
the conditions for the stability of periodic and quasi-periodic orbits
under small perturbations.  We point out that the KAM theorem was
derived for orbit structures under an applied potential, and not for 
self-organized collective instabilities.  Or in other words, for 
self-organized collective processes, the perturbations cannot be regarded 
as small from the outset of the growth of the instability.}, the multi-faceted 
correlations among the entire collection of N-particles are being established
to drive the emergence, maintenance and long-term evolution
of the collective instability. An analogy from a different
field is that of the growth of a living organism.  Most of us
would agree that if we simply assemble together
the material content of a living organism, no matter how closely
we can approximate the actual material makeup of the organism, we will 
never be able to ``breathe life'' into such an assembly of parts. 
The ``life force'' of a living being is derived from the very
self-organization process (i.e., the growth process of the biological
form), through the multitude of correlations established during
the ``co-evo" process.  Similarly for galaxies possessing a large-scale
density wave mode: the actual particles' trajectories supporting
the pattern acquire global coherence during the spontaneous growth
process of the mode, despite their chaotic appearance when each is
viewed individually. In some sense, the apparently ``ordered'' forms of the
periodic orbits for galactic systems are static and inert, and as a result
they lack the very life force (i.e., the ability to maintain themselves
and simultaneously to be able to evolve through the enabling roles
of self-organization and collective dissipation) we are talking about; 
whereas the apparently more ``chaotic'' looking trajectories in the N-body 
systems in fact contain innate orders and capacity for slow transformation
derived from the self-organization process, which facilitate
the further maintenance and evolution of the parent galactic system.
The equilibrium in physical galaxies is a dynamical equilibrium,
maintained as a balance of the growth and dissipation tendencies
of the wave mode (Zhang 1998), rather than the static equilibrium 
described by the passive orbit analysis and model-building approaches.

The analogies we made above between a biological system possessing living 
organisms and a galactic system possessing density wave modes are in fact not
accidental: both are instances of ``dissipative structures'' governed
by the laws of self-organization process for systems that
are open and far-from-equilibrium (Zhang 1998; Glansdorff \& Prigogine 
1971).

\subsection{Implications for the Kinematic and Dynamical States
of Nearby Galaxies}

Spiral and barred galaxies are expected to be systems
in near dynamical equilibrium, and thus their kinematic and 
dynamical states are partly reflected in their morphological 
appearances: this connection between morphology and kinematics
for quasi-equilibrium galaxies is the foundation of our approach for
corotation determination.  

So far, we have detected the phase shifts 
and derived corotation radii using $H$-band (1.65$\mu$m) images for 
more than 100 galaxies in the OSUBGS sample, which includes spiral
and barred galaxies of all morphological types, as well as using 
several galaxy images from other ground-based and space-based datasets.
Our initial success in applying this method, as reflected in
the observed correspondence between the corotation radii determined
by the phase shift method and the resonance features present in the 
original images, and the good agreement between our results and
many of the previously-published ones using other methods, indirectly confirms
that a large fraction of the observed disk galaxies in the nearby
universe have acquired a quasi-steady state with respect
to the formation of spiral/bar modes, since only in the quasi-steady
state of the wave mode can one expect a correlation between
the kinematic characteristics (such as corotation radius)
and the pure morphological appearance (i.e., the NIR image).

Over the past four decades since the advent of density
wave theory, the issue of whether the observed density wave
patterns in galaxies are quasi-stationary or else are transient
has always been a matter of debate (see, e.g. Binney \& Tremaine
1987 \S 6.4 and the references therein; Lucentini 2002).  Among the N-body 
simulated galaxies, results also differ between the quasi-steady patterns
obtained (Donner \& Thomasson 1994; Zhang 1998), and the more transient
patterns obtained (Sellwood \& Carlberg 1984; Sellwood \& Binney
2002).  We point out here that at least one of the known
causes of the differing results in these N-body simulations is
{\em the differing basic states of the galactic disks being used}.
When simulating spiral patterns, Sellwood \& Evans (2001)
and Sellwood \& Binney (2002) used a disk model that is over-stable
with respect to the formation of large-scale spiral modes. Therefore
any spiral perturbations will tend to die out, which is the reason
they only obtained transient spiral patterns in these studies.
However, growing observational evidence, including our
current work, shows that the large-scale spiral and barred
patterns in physical galaxies are in fact {\em 
gravitationally-unstable global modes}, not the transient structures
studied in the above work which used over-stable basic states.
As a matter of fact,  even
in some of the work conducted by these authors themselves
(i.e. Sparke \& Sellwood 1987; Debattista \& Sellwood 2000),
when a basic state model that allowed a genuine global
mode to emerge was used (in this case mostly bar modes and bar-driven spirals), 
they had invariably observed the same kind of collective effects
(including secular orbital change) and long-lasting patterns.

One other consequence of the presence of a phase shift, as we have already seen, 
is the secular evolution of the basic-state mass distribution for a galaxy 
containing large-scale, skewed density wave patterns.  
The secular morphological transformation mechanisms
proposed in the past involve mostly gas accretion
under a barred potential (Combes \& Sanders 1981;
Kormendy \& Kennicutt 2004 and the references therein), 
which were found to be inadequate to
transform the Hubble types of galaxies except for the very latest
types (Sc, Sd), due to the paucity of gas in most disk galaxies.
When the role of stars is also considered (Zhang 1996, 1998, 1999),
a more coherent and theoretically-sound basis for the
secular evolution paradigm emerges that is powerful enough to account
for both the morphological transformation of all galaxy types along
the Hubble sequence, and for the differing evolution
rate observed for galaxies residing in the field
and the cluster environments (cluster galaxies evolve faster because
of the large-amplitude, open spiral patterns excited
during the tidal interactions with neighboring galaxies
and with the cluster potential, and the secular evolution
rate is effectively proportional to the wave amplitude squared
and pattern pitch angle squared [Zhang 1998,1999]).

In general, the radial mass accretion/excretion rate across
any galactic radius can be calculated from (Zhang 1998)
\begin{equation}
{\overline{ {{dM} \over {dt}}}} (r)
=  - { {R} \over {v_c} }
\int_0^{2  \pi}
\Sigma(r,\phi)
{ {\partial {{{\cal V}}(r,\phi)}} \over {\partial \phi}}
d \phi
,
\label{eq:eqmass}
\end{equation}
where the surface density $\Sigma$ and potential ${\cal V}$ have
been written in their non-perturbative form (no subscript 1
as in the previous equations), since in this calculation we can
use the over-density and potential obtained directly from NIR
images.  But when the actual differentiation and integration are
carried out, only the perturbative quantities contribute
to the final value obtained (Zhang 1996).

\begin{figure}[ht]
\vspace{3.in}
\centerline{
\includegraphics{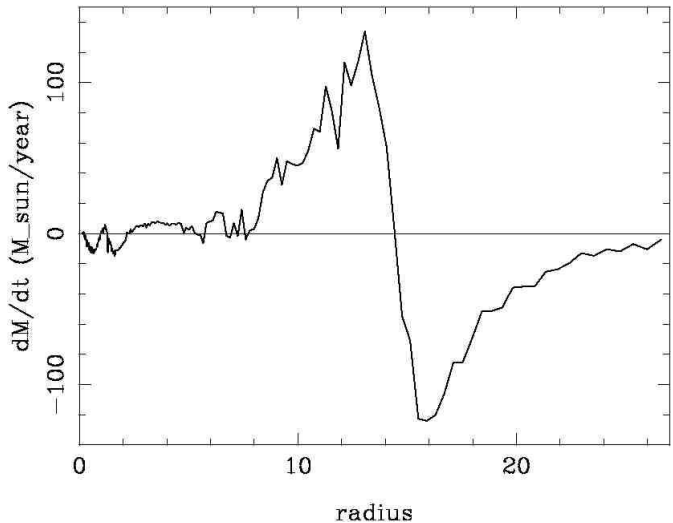}
}
\caption{Radial mass accretion/excretion rate calculated for
NGC 1530 from a $K_s$-band image (\S 3.2).}
\label{fg:fg1530mass}
\end{figure}

In Figure \ref{fg:fg1530mass}, we give an example of such a calculation for
the SBb galaxy NGC 1530, using the surface density map for NGC 1530 described 
in \S 3.2.  The methodologies needed for the calculation of phase shift
and accretion rates are different. The phase shift,
being dimensionless, can be determined without
knowledge of the distance and with reasonable accuracy under the assumption
of a constant M/L ratio. However, mass accretion rates require the
knowledge of real M/L ratios, galaxy rotation velocities, and 
absolute lengths.  Note that the shape of the phase shift distribution
curve (Figure \ref{fg:fg1530}, Left) is different from the mass accretion rate 
distribution (Figure \ref{fg:fg1530mass}).
This is because the mass accretion rate depends not only on the
phase shift, but also on the surface mass density distribution,
the distribution of the density wave pertubation amplitude,
as well as on the galaxy rotation curve distribution. 
The peak accretion rate for this galaxy is close to 140 M$_{\odot}$yr$^{-1}$
near the end of the main bar.  The inner accretion rate we obtained
for this galaxy (5-10 M$_{\odot}$yr$^{-1}$) is consistent
with the gas accretion rate estimated by Regan et al. (1997,
see their Figure 2), which is in the range of 2-5  M$_{\odot}$yr$^{-1}$ for 
the inner 4 kpc, since our result includes both the stellar and gaseous 
accretions.  Also note that our results indicated both accretion and excretion
in the central region due to nested resonance structures, with the
exact locations of these resonances expected to evolve over
long time as the basic state itself evolves.
The large excretion in the outer disk is expected to contribute
to the build-up of an extended outer envelope usually observed
for disk galaxies and disky ellipticals.

The result of the mass accretion calculation above shows that for at
least a fraction of the galaxies such as NGC 1530
which have extremely strong and open density wave patterns, the secular
evolution  can be very rapid (i.e. a mass accretion rate of $10^{10}$ 
solar masses per $10^8$ years is possible), compared to a typical galaxy 
like our own Milky Way where it takes about $10^{10}$ years to accumulate
$6 \times 10^{9}$ solar mass through the same mechanism (Zhang 1999), since
the Galactic spiral is both less intense and also more tightly wound.
We have so far calculated the mass accretion rates for several
other galaxies as well, and found that a peak accretion rate
of several to several tens of solar masses per year is typical of galaxies
of intermediate Hubble types.  An associated heating effect
as a by-product of the energy and angular momentum exchange
process between the density wave and the basic-state stars
results also in the secular heating of the disk stars
in addition to their secular orbit migration (Zhang 1999),
and this results in the vertical migration of the stellar
trajectories as well as the radial migration, which together lead
to the building up of the bulge star populations (Zhang 2003).
We note here that some of the past-proposed bulge-building
mechanisms involving vertical resonances (e.g. Pfenniger \& Friedli 1991)
are not likely to be effective since the single-orbit resonance
condition is likely to be destroyed by the presence of collective
instabilities, as our earlier analysis had shown.  In fact, the
so-called ``buckling instability'' attributed to those vertical
resonances acts very fast (Pfenniger \& Friedli 1991), and would have 
resulted in two distinct classes of disk galaxies (one before the buckling,
with no bulge, and one after the buckling, with bulge), whereas
the observed bulges come in all variations of bulge-to-disk
ratio, and there is no signature of the above bi-modality which
would have resulted if buckling is the underlying mechanism 
for bulge formation.  The buckling observed in previous 3D
N-body simulations are apparently a result of the initial
conditions chosen in these simulations, which results in the
rapid evolution of the initial configuration to reach a 
new equilibrium quasi-steady state.

Note that past N-body simulations have had difficulty reproducing
a large accretion rate required for the cosmological
transformation of the Hubble types of galaxies
(e.g. Zhang 1998, 1999).  This is now understood to be mainly due to
the fact these simulations used a much smaller number of particles 
than the number of stars present in physical galaxies ($10^6$ rather
than $10^{11}$).  This produced an increase in the initial growth
rate of bar modes, as well as in the collisional
relaxation induced by the collective instability, which
together demand a rigid halo/bulge component to be imposed 
(especially for the 2D N-body simulations) in order to artificially
suppress the violently growing bar modes, so that its
resulting rapid heating would not suppress the appearance
of the spiral modes (Ostriker \& Peebles 1973).  
However, the suppression of the density modes
with large growth rate also means that in these simulations we
have only simulated the class of spiral patterns which have
small perturbation amplitudes (the density perturbation amplitudes
in the simulated spirals are usually around 15-20\%, and the
potential perturbation amplitudes are around 5\%).  The small
wave amplitude in the simulated spirals is the main
reason for the small accretion rate observed in N-body
simulations, a conclusion already reached in Zhang (2003).  
These facts reflect the inherent difficulty in using small number of
particles to simulate a highly nonlinear structure which has large
perturbation amplitude.  Physical galaxies
have significantly more particles than
used in the simulations so far and thus can sustain
sufficiently large arm/inter-arm contrast for their density waves 
to result in the kind of secular 
evolution rates observed for cluster and field galaxies. 

We also comment here that in the past other researchers had
attempted to use {\em resonance broadening} of a time-varying
density wave pattern to reconcile
the observed secular evolution behavior of particle
trajectories found in N-body simulations with the LBK conclusion
of no-wave-basic-state-interaction-except-at-resonances
(see, e.g., Sellwood \& Binney 2002 and the references therein).
We point out that the secular evolution process discussed in
this work is not a direct result of the time-fluctuating
nature of the density wave: in fact, the mass accretion
rate expression (equation \ref{eq:eqmass}) becomes exact
precisely at the quasi-steady state of the wave mode (Zhang 1998).
Indeed there are microscopic time fluctuations of the
overall potential in any N-particle disk, but such microscopic
fluctuations are not an integral part of the resonance-broadening
analysis (i.e., one could use different numbers of particles,
different initial conditions, etc. in these simulations, and still obtain the 
same mode and the same secular evolution rate, as long as the basic state is 
chosen to be the same.  The only change with the use of different particle 
numbers is the resulting heating rate.  But the mass-accretion rate 
and angular-momentum-exchange rate remain the same. See Zhang
1998 \S 5.2 for further details of these tests).
Another indication that the secular evolution and collective
dissipation behavior we discussed is unrelated to the broadening
of resonances is that, as recently found by Sellwood \& Binney
(2002), the resonance-broadening
approach produced radial migrations of particles, in an over-stable
disk subject to transient spiral perturbations, 
that were {\em exactly opposite} in direction
to what were predicted in the secular evolution theory, i.e. to
what was also observed in N-body simulations containing
unstable global modes.  The resonance-broadening
arguments thus clearly failed to capture the essential physics
of the secular evolution process in physical galaxies containing
global density wave modes.

\subsection{Phase Shifts, Torques, and Angular Momentum Exchanges}

In this subsection we address several issues related to
the phase shifts, torques, as well as angular momentum transports
and exchanges in disk galaxies, which underlies the approach we used
for determining the corotation radii and for calculating the mass
accretion rates.

There are actually two different meanings when we use the word torque 
in the context of spiral waves. There is the torque expression (1)
which (when multiplied by $2 \pi r$) is the total torque experienced by
the matter within a unit-width annulus due to the torquing
of the wave (we had called this $T(r)$ in Zhang [1996], or
$- T_1(r)$ as defined in Zhang [1998]. Hhere the negative sign in front
of $T_1(r)$ signifies the fact that $T_1(r)$ represents rather the
torquing in the $-\phi$ direction that
the wave applied onto the disk matter inside corotation),
which is effectively due to the rest of the disk
matter not in the annulus; there is also the torque
coupling integral that LBK had used in their 1972 
paper (which can be further broken up into the gravitational torque 
coupling integral $C_g (r)$ and advective torque coupling integral $C_a (r)$), 
which expressed the torquing of 
the inner disk on the outer disk, or vice versa, across a circular
boundary at a given radius $r$ of the disk. 
The {\em radial gradient} of the LBK's gravitational torque coupling 
integral $dC_g/dr$ is equal to our torque $T_1(r)$ in the linear regime
of the wave mode (here the sign convention works such that the positive
$dC/dr$ which signifies an angular mommentum drain from an annulus
is related to the $- \phi$ direction of the torque of the wave
on the disk matter).  At the quasi-steady state of the wave mode it is 
rather the {\em radial gradient of the sum} of the LBK's 
gravitational and advective torque coupling integrals 
that equals to the torque expression $T_1(r)$ (Zhang 1998, 1999).
Incidentally, the LBK total torque coupling integral $C(r) = C_a(r)+C_g(r)$
is the same as the total outward angular momentum
flux across a given radius r, if we are dealing
with spiral density wave.

The significance of these relations between the LBK torque coupling
integrals and our torque integral $T_1(r)$ is that the LBK's total torque 
coupling $C(r)$ needs to be of a bell shape in order for the relevant modes 
to be able to spontaneously and homogeneously emerge from an originally 
featureless disk, and this bell-shaped torque coupling corresponds to
a radial distribution of $T_1(r)$ (which is itself proportional 
to $\sin (m \phi_0)$, with $\phi_0$ the potential-density phase shift)
which has a positive
hump inside corotation, and negative hump outside corotation, exactly 
the same as the distribution of the phase shift we have determined 
in this work.  The change of sign of phase shift at corotation
can be shown to be a direct consequence of a mode's capability to 
spontaneously emerge out of an initially axisummetric, differentially rotating 
disk.  This is because in a differentially rotating disk, 
the density wave mode formed has negative angular
momentum density (with respect to the axisymmetric state of the disk) inside
corotation, and positive angular momentum density outside corotation.  In 
order for the wave mode to be able to spontaneously emerge, angular momentum 
had to be taken out from every annulus of the inner disk inside $r_{CR}$, 
and deposited onto every annulus of the outer disk outside $r_{CR}$ (and unlike 
what LBK had originally thought, that this taken-away and deposition 
happen only at resonances).  This fact 
also explains why the spontaneously-growing type of modes had to be of spiral 
shape (even for the bar mode, when it first emergies from an N-body disk, 
had been invariably shown to be of spiral shape initially
and then gradually evolves into straight bar.  See for example
the simulation of Sparke \& Sellwood [1987]): because a perfectly straight 
bar will not be able to transport angular momentum outward, and without 
lowering the angular momentum density in the inner disk (up to the 
corotation radius of the mode), a mode would not be able to emerge.

The above refers mostly to the linear modal-growth regime. Then, at the
quasi-steady state of the wave mode, since by definition the wave amplitude 
can no longer grow, yet the bell-shaped torque coupling remains as long
as the mode still contains some degree of skewness,
and thus the angular momentum is being continuously transported outward
(as well as being taken out from every annulus from the inner disk, and 
deposited onto every annulus on the outer disk), the out-transported
angular momentum cannot come from the wave itself, and has to come from the
basic state: therefore the basic state evolution (i.e., matter 
inflow and outflow) is a natural outcome of the combined facts of
the quasi-stationary nature of the wave mode and the continued outward
angular momentum transport by the wave mode at its quasi-steady state.  

Such a wave/basic-state interaction at the quasi-steady 
state of the wave mode is not due to the interaction of the mean wave 
potential with a single orbit (since the conservation of the Jacobi integral 
in the corotating frame of the mode forbids the single orbit to lose energy 
and angular momentum), but has to be through the collective instability at 
the spiral arms, with its integral expression given by equation 1.  
The effect of the torque integral $T(r)$ on a group of stars 
had been quantitatively confirmed through N-body simulations
(Zhang 1998, \S5.2.2, see especially the text in relation to Figure 10
in that paper).  

As regard to the ``solid body'' appearance of the 
wave/basic-state interaction described by equation 1,
despite the seeming randomness of each
star's orbital motion with respect to the wave, 
we want to point out the difference of torque action in three types of 
scenarios: (a) A true solid body as in the case of a curved sword,
or a scimitar (an analogy made by B. Elmegreen 2006, private
communication). Here we have the gravitational torque, and torque couple
(but no advective torque couple, since the particle configuration is
not in motion), yet no angular momentum transport. This is because for 
every segment of the sword the torque coupling integrals from the inner 
and outer boundaries of the segment exactly cancel each (the gravitational 
torque couple still has a gradient as in the case of a spiral mode, but this 
gradient is compensated by the countering torque gradient due to the
molecular bounding forces). That we can have torque and counter-torque but
no net angular momentum exchange in this situation is similar to one's
pushing a brick wall: one can apply a force, and experience a counter force, 
but unless the wall moves one will not have done any work on the wall. 
(b) A loose association of self-gravitating particles arranged in the shape 
of a curved sword.  In this case, there will be gravitational torque as before 
(and there might even be advective torque when the particles start moving), but 
the configuration will change as a result, and there is indeed angular 
momentum transfer between the inner and outer portions of this loose 
association of particles (they will swing in opposite directions to 
straighten out the curve of the sword, yet the total angular momentum 
of the system remains zero).  (c) The case of spiral modes in galaxies. This
situation is unlike that of either (a) or (b), in that there is 
simultaneously the gravitation/advective torque and torque couple, the 
transfer of angular momentum, as well as the maintenance of the coherence 
of the original modal configuration.  This structural stability of the mode
in the face of perturbing influences is what made the global instability
patterns so unique.  The structural stability is achieved as a dynamical 
balance between the spontaneous growth tendency of the unstable mode, and 
the dissipative damping tendency of the collective viscous forces. 
The wave acts as if it is a solid entity which can receive and give-out 
energy and angular momentum, yet it simultaneously possesses a
soft boundary in the sense that the constituting stars which support the 
wave motion swarm in and out of the wave crest.

\subsection{Physical Basis Underlying the Validity and Accuracy
of the Phase Shift Method}

The validity of the phase shift method for corotation determination
is rooted in the global self-consistency requirement 
of the wave mode (i.e., both the Poisson equation and
the equations of motion need to be satisfied at the same time). 
Therefore, the appearance that the corotation can be determined
from the Poisson equation alone, without knowledge of the kinematic
state of the disk matter, is only an illusion. In actuality
the global self-consistency between the Poisson equation and the
equations of motion has been enforced by nature itself when the
galactic resonant cavity filters and selects the set of modes it can support.
The phase shift resulting from the Poisson equation is not only determined by 
the spiral pitch angle, but also by the radial density variations 
of the non-axisymmetric density wave features (Zhang 1996).  
For a self-sustained global spiral mode, the radial density variation 
of the modal perturbation density as well as the pitch angle variation 
together would be such that the Poisson equation will
lead to the zero crossing of the phase shift curve to be
exactly at the corotation radius of the mode, since that's where
the peak of the bell-shaped angular momentum flux is, and the
radial gradient of the angular momentum flux is proportional
to the sine of the phase shift (Zhang 1998; see also
Lin \& Lau 1979, equation [C8]).  Since the bell-shaped angular
momentum flux distribution is what allowed the spontaneous growth
of the mode in the linear regime,
only modes with density distributions which result in the correct
phase shift distribitions are naturally selected by the galactic
resoanant cavity.

The self-consistency requirement of a spontaneously-formed density
wave mode explains not only why we can use the Poisson equation alone
to determine the corotation radius, but also why we can use only continuity
equation alone to determine the pattern speed as in the TW approach: each
is only seeing one side of the coin, but since both sides join at a unique
boundary, we can determine the circumference of the coin from
measurements conducted on either side.

The phase shift distribution calculated through the Poisson equation
would be the same whether the pattern is rotating or not, as long
as we used the same surface density distribution. 
The phase shift method is thus inherently NOT a rotation detection method. 
Rather, it is a diagnostic tool for characterizing the features of self-organized 
modes in galactic disks.  We can imagine, for the sake
of the argument, that a mode had initially reached quasi-steady state 
with a given pattern speed. If, suddenly, we apply a forced potential 
of exactly the same shape as the original modal potential, but with
a somewhat faster pattern speed than the original natural
pattern speed, the pattern density will tend to follow this applied
potential's forcing with only minor changes in shape.  The phase shift method
would not be able to predict that the rotation speed of the pattern
has changed in this case, or equivalently the corotation radius has changed.  
But once the forcing is removed, the pattern will quickly recover to its 
original pattern speed. The reason is that the faster pattern speed had moved 
the corotation radius from its original, natural position, which 
previously allowed the sign of the 
angular momentum deposition by the wave everywhere across the disk to match 
the sign of the angular momentum density of the wave mode itself.  
This matching condition 
is violated as in the forced case, once we remove the forcing, the faster mode
will be damped out because of the wrong sign and wrong amounts
of angular momentum being deposited by the wave mode over much of its extent.
The original mode will re-appear since it is naturally amplified 
in the galactic resonant cavity determined by the basic state.
Or to put it in another way, the spiral shock and the 
associated collective dissipation
tends to destroy or damp the wave at every instant of the time.  It is
only through the continued amplification tendency of the original mode in the
galactic resonant cavity that a quasi-steady state can be maintained.
{\em The modes that satisfy the self-sustainability condition have a
special appearance that matches its kinematic characteristics, so that
the growth tendency balances the dissipation tendency to reach
a dynamical equilibrium at the quasi-steady state}. 

We thus see that as a general method for corotation determination 
the visual matching of the forced-pattern response with that of the
observed galaxy morphology is not likely to be reliable, since the 
response pattern tends to look almost identical for a quite wide
range of pattern speed.  The limitations of the passive simulation
approach becomes all the more prominent when we realize that realistic
galaxies contain nested density wave patterns which all rotate at
different pattern speeds, as we have found out in this work.

On the other hand, the phase shift method itself
is only expected to be reliable for cases of
spontaneously formed density wave patterns that have reached
quasi-steady state.  We have found empirically from analysing the
OSUBGS that usually the early-type disk galaxies gave the best fit
between the predicted corotations and the resonance features,
likely related to the fact that early type galaxies have had the
longest time to evolve and reach quasi-steady state. The latest
Hubble types usually have more flocculent patterns, and the phase
shift result correspondingly does not show a well-organized pattern,
reflecting partly the fact that the wave-mode pattern speeds and corotation
radii in these galaxies are themselves poorly defined.
The intermediate Hubble types have strong and
open spiral arms, but the results of the phase shift calculation
tend to be more sensitive to the uncertainty in inclination-angle
deprojections which can result in inaccurate spiral pitch angle 
representations, an issue we will further address in the sequel paper.
The intrinsic accuracy of the phase shift method should be 
high if we are dealing with an ideal, spontaneously-formed mode
that has reached quasi-steady state, and if we have accurate knowledge of
of the orientation parameters of the galaxy.  
In practice, however, the mode may not be in quasi-steady state
and/or the orientation parameters may be uncertain.  Since the errors will
likely to be dominated by unknown systematics rather than by
random measurement errors, we have not assigned error bars to
our results. Other sensitivities/pitfalls of the method 
will be discovered as it is applied to a greater number of
real and simulated galaxies.

We now comment briefly on the pattern speeds and corotation radii
determined from previous modal calculations and their relation to
the results of our current study.
In the 1970s and 1980s Lin and collaborators constructed 
self-consistent spiral modes both through solving
the exact fluid set of equations in the linear regime, as well as through 
the asymptotic approach of higher-order local calculations which 
obtained piece-wise continuous wave trains, and then joined them together 
to derive the quantum conditions for the pattern speed determination
(see, e.g. Lin \& Lau 1979 and the 
references therein; Bertin et al.  1989a,b and the references therein).
These two approaches were found to agree with one another
as long as one deals with slow-growing modes.
In both approaches, the quantum condition for the mode
was derived by enforcing the global 
self-consistency requirement between the Poission equations and the 
equations of motion (or the 
so-called Poisson-Euler set in the fluid approach).  
In the asymptotic approach, this global self-consistency is 
acquired in a piece-wise, locally-self-consistent manner, and by throwing away 
the so-called ``out of phase'' terms (which exactly
corresponds to the potential-density phase shift we had discussed
in this work).  The fact that this
latter approach leads to results similar to the ``exact'' 
numerical solutions which accurately contain all the phase shift 
information is because the asymptotic approach only solved the 
weakly-growing kind of modes (i.e., either very small pitch angles,
as in tightly-wound early Hubble type galaxies, or very large pitch angles, 
as in very straight bars, again appearing in early Hubble type galaxies;
both of these cases give small phase shifts, since the phase shift is 
largest for pitch angle close to 45 degrees, and is zero for 
pitch angles of 0 and 90 degrees).  Therefore, the asymptotic approach worked 
precisely for situations where the phase shift is small, so
the resulting small ``out-of-phase'' terms (which reveal the effect
of potential-density phase shift) can be thrown away.
For the fast growing ``violently unstable'' modes which have a more
open spiral morphology (which resemble intermediate Hubble type galaxies), 
the exact numerical solutions of linear global modes do contain the phase shift 
distribution that changes sign at corotation (C.C. Lin 1995, private
communication), and it is precisely this phase shift distribution
in the open spiral modes which leads to the large
growth rate of the mode in the linear regime, and to the rapid
azimuthal steepening of the sinusoidal wave profile into narrow
shock fronts in the nonlinear regime (Zhang 1998). 

Realistic galaxies also often contain companions and/or have
gone through major/minor mergers.  We will explore in future
studies the role of environment on the reliability and the
detailed distribution of phase shifts in galaxies.  We comment here that
if the galaxy disks involved are unstable to the formation of
intrinsic modes, the effect of an interaction often only
serves to excite a large-amplitude mode which has the same
properties as the intrinsic mode (Zhang et al. 1993),  i.e., the intrinsic 
modes in unstable disks can be rather robust during modest galaxy
interactions.  Minor mergers potentially can result in a change
of the basic state, and thus the kind of unstable modes admitted. The phase 
shift method should be applicable in this case as soon as the entire 
configuration has settled into the new quasi-steady state.  Major mergers,
on the other hand, often result in damages that are
so severe that the galaxy morphology is no longer described
by grand-design density wave patterns, so the phase shift method
will not apply. For certain fortuitous cases such as NGC 4622 the method
could still work even for transient waves, but in general, its 
validity has to be checked on an individual basis for any galaxy
involved in interactions.

\section{CONCLUSIONS}

In this work, we have developed and verified a general approach for
determining the corotation radii of density waves in disk galaxies,
which applies to any disk galaxies whose density
wave modes have reached quasi-steady state -- a condition empirically
found to be the case for the majority of the nearby disk galaxies.  

Compared to other approaches developed so far,
the new approach for corotation determination has a number of
advantages: (1) It is relatively insensitive to star formation, 
M/L variations, and vertical scale height assumptions, at least at the 
level these problems are present in the NIR.  (2) It can be applied to 
face-on galaxies; the orientation of the bar has little impact. 
(3) It can be applied effectively to all Hubble types, at
least those with a disk shape. (4) The radial extents of the nested 
resonances and multiple pattern speeds can be determined in a 
model-independent fashion, and in many cases the central resonances
can be determined with ``super-resolution'' due to the non-local 
nature of the potential calculation.  (5) The application of the method 
can use existing databases of NIR galaxy images without the need for 
significant investments in new telescope time.
Furthermore, the simplicity of the application of the method allows the
statistical analysis of the correlation of the calculated
corotation radii with the observed morphological features of galaxies.
(6) The connection between the potential-density phase shifts and the
secular evolution rates of galaxies facilitates the study of the galaxy
morphological and kinematic features in the context of the cosmic
evolution of galaxies. This provides another way of quantifying the 
impact of bars and spirals on disk galaxies.

\section*{ACKNOWLEDGMENTS}

We thank E. Laurikainen for some of
the deprojected images used in this study, and E. Athanassoula,
G. Contopoulos, B. Elmegreen, J. Fischer, P. Grosbol,
D. Pfenniger, H. Salo, S. Tremaine as well as an anonymous
referee for helpful comments and exchanges which
improved the presentation of the text.
XZ acknowledges the support of the Office of Naval Research.
RB acknowledges the support of NSF grant 
AST 050-7140 to the University of Alabama. 
Funding for the Ohio State University Bright 
Galaxy Survey was provided by NSF Grants AST 
92-17716 and AST 96-17006, with additional funding 
from the Ohio State University.  The NASA/IPAC Extragalactic Database 
(NED)
is operated by the Jet Propulsion Laboratory, California Institute of
Technology, under contract with NASA. 

\section*{REFERENCES}

\noindent
Arsenault, R., Boulesteix, J., Georgelin, Y.,
\& Roy, J.R. 1988, A\&A, 200, 29

\noindent
Athanassoula, E. 2003, MNRAS, 341, 1179

\noindent Abraham, R.G., Merrifield, M.R., Ellis, R.S.,
Tanvir, N.R., \& Brinchmann, J. 1999, MNRAS, 308, 569

\noindent
Ball, R. 1992, ApJ, 359, 418

\noindent
Bell, E.F. \& de Jong, R.S. 2001, ApJ 550, 212.

\noindent
Bertin, G., Lin, C.C., Lowe, S.A., \& Thurstans, R.P. 1989a,
ApJ, 338, 78; 1989b, ApJ, 338, 104

\noindent Binney, J., \& Tremaine, S. 1987, Galactic Dynamics
(Princeton:Princeton Univ. Press)

\noindent
Block, D.L., Buta, R.J., Knapen, J.H., Elmegreen, D.M.,
Elmegreen, B.G., \& Puerari, I.P. 2004, AJ, 128, 183

\noindent
Buta, R. \& Purcell, G. B. 1998, AJ, 115, 494

\noindent
Buta, R., Alpert, A., Cobb, M. L., Crocker, D. A., \& Purcell, G. B.
1998, AJ 116, 1142

\noindent
Buta, R.J., Byrd, G., \& Freeman, T 2003, AJ, 125, 634

\noindent
Buta, R.J. Mitra, S., de Vaucouleurs, G., \& Corwin, H.C.
1994, AJ 107, 118

\noindent
Buta, R.J., Vasylyev, S., Salo, H.
\& Laurikainen, E. 2005, AJ 130, 506

\noindent
Buta, R.J., \& Zhang, X. 2007, in preparation

\noindent Butcher, H., \& Oemler, A. Jr. 1978, ApJ, 219, 18; 
ApJ, 226, 559

\noindent
Byrd, G., Freeman, T., Howard, S., \& Buta, R. 2006,
submitted to AJ

\noindent
Canzian, B. 1993, ApJ, 414, 487

\noindent
Canzian, B. 1998, ApJ, 502, 582

\noindent
Clemens, M.S., \& Alexander, P. 2001, MNRAS, 321, 103

\noindent
Combes, F. \& Sanders, R.H. 1981, A\&A, 96, 164

\noindent
Contopoulos, G. 1980, A\&A, 31, 198

\noindent
Contopoulos, G. \& Grosbol, P. 1986, A\&A, 155, 11

\noindent
Contopoulos, G. \& Grosbol, P. 1988, A\&A, 197, 83

\noindent
Contopoulos, G. \& Mertzanides, C. 1977, A\&A, 61, 477

\noindent
Contopoulos, G. \& Voglis, N. 1996, in Barred Galaxies,
ASP Conf. Series, vol. 91, 321, eds. R. Buta, D.A. Crocker, \& B.G.
Elmegreen

\noindent Couch, W. J., Ellis, R. S., Sharples, R. M., \& Smail, I. 
1994,
ApJ, 430, 121

\noindent
de Grijs, R. 1998, MNRAS, 299, 595

\noindent
de Vaucouleurs, G. 1963, ApJS, 8, 31

\noindent
de Vaucouleurs, G., de Vaucouleurs, A., Corwin, H. G., Buta, R.J.,
Paturel, G., \& Fouq\'ue, P. 1991, Third Reference Catalogue of Bright
Galaxies (New York, Springer)

\noindent
Debattista, V.P., \& Sellwood, J.A. 2000, ApJ, 543, 704

\noindent
Donner, K. J. \& Thomasson, M. 1994, A\&A, 290, 785

\noindent
Dressler, A. 1980, ApJ, 236, 351

\noindent Dressler, A., Oemler, Jr. A., Butcher, H.R., \& Gunn, J.E. 
1994, ApJ, 430, 107

\noindent Egusa, F., Sofue, Y., and Nakanishi, H. 2006, 
astro-ph/0410469

\noindent
Ellis, R.S. 1997, ARA\&A, 35, 389

\noindent
Elmegreen, B. G., \& Elmegreen, D. M. 1983, ApJ, 267, 31

\noindent
-----. 1985, ApJ, 288, 438

\noindent
------. 1989, in Evolutionary Phenomena in Galaxies,
ed. J.E. Beckman \& B.E.J. Pagel (Cambridge: Cambridge University 
Press),
p.83.

\noindent
Elmegreen, B.G., Elmegreen, D.M., \& Seiden, P. E.  1989, ApJ, 343, 602

\noindent
Elmegreen, B.G., Wilcots, E., \& Pisano, D.J. 1998, ApJ, 494, 37

\noindent
England, M.N. 1989, ApJ, 344, 669

\noindent
Erwin, P., \& Sparke, L.S. 1999, in Galaxy Dynamics, ASP Conf. Ser.
vol 182 (eds. D. Merritt \& J.A. Sellwood), p. 243

\noindent
Eskridge, P.B. et al. 2002, ApJS, 143, 73

\noindent
Friedli, D. \& Martinet, L. 1993, A\&A,
277,27

\noindent
Frogel, J.A., Quillen, A.C., \& Pogge, R.W. 1996, in
New extragalactic perspectives in the New South Africa, 
eds. D. L. Block and J. M. Greenberg (Dordrecht: Kluwer),
Ap\&SS 209, 65

\noindent
Gebhardt et al. 2000, ApJ, 539, L13

\noindent
Gerssen, J., Kuijken, K., \& Merrifield, M.R. 1999, MNRAS, 306, 926

\noindent
Glansdorff, P., \& Prigogine, I. 1971, Thermodynamic Theory of
Structure Stability and Fluctuations (New York: Wiley-Interscience)

\noindent
Hernandez, O. et al. 2005, ApJ, 632, 253

\noindent
Hunter, J.H., England, M.N., Gottesman, S.T.,
Ball, R., \& Huntley, J.M. 1988, ApJ, 324, 721

\noindent
Kenney, J.D.P., Wilson, C.D., Scoville, N.Z., Devereux, N.A., Young, 
J.S.
1992, ApJL, 395, 79

\noindent
Kennicutt, R.C. et al. 2003, PASP, 115, 928

\noindent
Koo, D.C., \& Kron, R.G. 1992, ARA\&A, 30, 613

\noindent
Kormendy, D. \& Kennicutt, R. 2004, ARA\&A, 
42, 603

\noindent
Laine, S., Shlosman, I., Knapen, J.H., \& Peletier, R.F. 2002,
ApJ, 567, 97

\noindent
Laurikainen, E., Salo, H., Buta, R.J., \& Corwin, H.C. 1994, AJ,
107, 118

\noindent
Laurikainen, E., Salo, H., \& Buta, R. 2005, MNRAS, 362, 1319

\noindent
Laurikainen, E., Salo, H., \& Rautiainen, P. 2005, MNRAS, 331, 880

\noindent
Lilly, S., Abraham, R., Brinchmann, J., Colless, M., Crampton, D.,
Ellis, R., Glazebrook, K., Hammer, F., Le Fevre, O., Mallen-Ornelas, 
G.,
Shade, D. \& Tresse, L. 1998, in The Hubble Deep Field, eds.
M. Livio, S.M. Fall, \& P. Madau (Cambridge: CUP)

\noindent
Lin, C.C. 1970, in IAU Symp. 38, The Spiral Structure of Our Galaxy,
eds. W. Becker \& G.I. Contopoulos (Dordrecht: Reidel), 377

\noindent
Lin, C.C., \& Lau, Y.Y. 1979, Studies in Appl. Math., 60, 97

\noindent
Lin, C. C., \& Shu, F. H. 1964, ApJ, 140, 646

\noindent
Lindblad, P.A.B., \& Kristen, H. 1996, A\&A, 313, 733

\noindent
Lindblad, P.A.B., Lindblad, P.O., \& Athanassoula, E. 1996
A\&A, 313, 65

\noindent
Lucentini, J. 2002, Sky \& Telescope, September, p.36

\noindent
Lynden-Bell, D., \& Kalnajs, A.J., 1972, MNRAS, 157, 1

\noindent
Mark, J.W.-K. 1976, ApJ, 205, 363

\noindent
Merritt, D. \& Ferrarese, L. 2001, ApJ, 547, 140

\noindent
Oey, M. S., Parker, J. S., Mikles, V. J., \& Zhang, X. 2003, 
\aj, 126, 2317

\noindent
Ostriker, J.P., \& Peebles, P.J.E. 1973, ApJ, 186, 467

\noindent
Pfenniger, D. 1986, A\&A, 165, 74

\noindent
Pfenniger, D., \& Friedli, D. 1991, A\&A, 252. 75

\noindent
Press, W.H., Teukolsky, S.A., Vettering, W.T., \& Flannery, B.P. 1992,
Numerical Recipes in Fortran (Cambridge: CUP)

\noindent
Puerari, I. and Dottori, H. 1997, ApJ, 476, L73

\noindent
Purcell, G. 1998, PhD Thesis, University of Alabama

\noindent
Quillen, A.C., Frogel, J.A., Gonz\'alez, R.A. 1994,
ApJ, 437, 162

\noindent
Rand, R.J. 1995, AJ 109, 2444

\noindent
Rautiainen, P. \& Salo, H. 1999, A\&A, 348, 737

\noindent
Rautiainen, P., Salo, H., \& Laurikainen, E. 2005, ApJ, 631, L129

\noindent
Regan, M.W., Vogel, S.N., \& Teuben, P.J. 1997, ApJL, 482, 143

\noindent
Regan, M.W., Teuben, P.J., \& Vogel, S.N. 1996,  AJ, 112, 2549

\noindent
Salo, H., Rautiainen, P., Buta, R., Purcell, G.B., Cobb, M.L.,
Crocker, D.A., \& Laurikainen, E. 1999, AJ, 117, 792

\noindent
Sandage, A. 1961, The Hubble Allas of Galaxies (Washington DC: 
Carnegie)

\noindent
Schwarz, M.P. 1984, MNRAS, 209, 93

\noindent
Schlegel, D. J., Finkbeiner, D. P., \& Davis, M. 1998, \aj, 500, 525

\noindent
Sellwood, J.A., \& Binney, J.J. 2002, MNRAS, 336, 785

\noindent
Sellwood, J.A., \& Evans, N.W. 2001, ApJ, 546, 176

\noindent
Shu, F.H., Stachnik, R.V., \& Yost, J.C. 1971, ApJ, 166, 465

\noindent
Sparke, L.S., \& Sellwood, J.A. 1987, MNRAS, 225, 653

\noindent
Tagger, M., Sygnet, J.F., Athanassoula, E., \& Pellat, R. 1987,
ApJ, 318, L43

\noindent
Toomre, A. 1981, in Structure and Dynamics of Normal Galaxies,
ed. S. M. Fall \&
D. Lynden-Bell (Cambridge: Cambridge Univ. Press), 111

\noindent
Tremaine, S., \& Weinberg, S. 1984, ApJ, 282, 5

\noindent
Tully, R. B. 1988, Nearby Galaxies Catalog (Cambridge: 
Cambridge Univ. Press)

\noindent
Voglis, N., Stavropoulos, I. \& Kalapotharakos, C. 2006, MNRAS on-line
(astro-ph/0606561)

\noindent
Weinberg, M.D. 1993, ApJ, 410, 543

\noindent
Worthey, G. 1994, ApJS, 95, 107

\noindent
Zhang, X. 1996, ApJ, 457, 125

\noindent
Zhang, X. 1998, ApJ, 499, 93

\noindent
Zhang, X. 1999, ApJ, 518, 613

\noindent
Zhang, X. 2003, Journal of Korean Astronomical Society, 36, 223

\noindent
Zhang, X., \& Buta, R.J. 2006, in Galaxy Evolution Across the Hubble
Time, Proc. IAU Symp. 235, eds F. Combes and J. Palous, in press

\noindent
Zhang, X., Wright, M., Alexander, P. 1993,
ApJ, 418, 100

\vfill
\eject

\end{document}